\documentclass[12pt]{article}
\usepackage{graphicx,amsmath,amssymb,slashed,hyperref,bbold,fullpage,cancel}
\usepackage[toc,page]{appendix}

\newcommand{\tr}{\text{Tr}}
\newcommand{\cpn}{$\mathbb{CP}(N)$ }
\newcommand{\cpnm}{$\mathbb{CP}(N-1)$ }

\newcommand{\flgn}{$\mathcal{F}_{\left\lbrace N_\alpha \right\rbrace}$}
\numberwithin{equation}{section}
\newcommand{\spn}{\text{span}}

\begin{document}
	\title{\begin{flushright}
			{\small FTPI-MINN-19/22, UMN-TH-3832/19 }
		\end{flushright}
		{\Large{\textbf{
					\rule{0mm}{15mm} General Composite Non-Abelian Strings and Flag Manifold Sigma Models
	}}}}

\author{Edwin Ireson$^{1,2}$\\
\textit{\small $^{1}$ School of Physics and Astronomy,}
\\ \textit{\small University of Minnesota, Minneapolis, 55455, USA }\\
\textit{\small $^{2}$ William I. Fine Theoretical Physics Institute,}
\\ \textit{\small University of Minnesota, Minneapolis, MN, 55455, USA}
	}
\maketitle

\begin{abstract}
We fully investigate the symmetry breaking patterns occurring upon creation of composite non-Abelian strings: vortex strings in non-Abelian theories where different sets of colours have different amounts of flux. After spontaneous symmetry breaking, there remains some internal colour degrees of freedom attached to these objects, which we argue must exist in a Flag manifold, a more general kind of projective space than both  \cpn and the Grassmannian manifold. These strings are expected to be BPS, since its constituents are. We demonstrate that this is true and construct a low-energy effective action for the fluctuations of the internal Flag moduli, which we then re-write it in two different ways for the dynamics of these degrees of freedom: a gauged linear sigma model with auxiliary fields and a non-linear sigma model with an explicit target space metric for the Flag Manifolds, both of which $\mathcal{N}=(2,2)$ supersymmetric. We finish by performing some groundwork analysis of the resulting theory.
\end{abstract}

\section{Introduction}
The \cpn non-linear sigma model has undergone much analysis in many contexts, in particular because it provides a very tractable (in its simplest formulation, exactly solvable) theory in which confinement occurs \cite{Witten:1978bc}. Being a K\"{a}hler manifold, it is then particularly straightforward to study supersymmetric enhancements thereof, and leads to a rich study of deformations by superpotentials and other very geometric considerations. 

They appear quite naturally when developing a worldsheet action for non-Abelian vortex strings \cite{HT1,ABEKY,SYmon,HT2},
(see \cite{Trev,Jrev,SYrev,Trev2} for reviews) including its heterotic versions \cite{3sv}. In four-dimensional theories with a $SU(N+1)$ gauge group and a scalar symmetry breaking potential, solitonic vortex string solutions can be constructed: certainly some exist that are merely copies of the usual Abelian Abrikosov-Nielsen-Olesen vortex, in which all of the non-Abelian gauge symmetry is completely broken, but more elementary strings (of lower tension) can be obtained by allowing some leftover invariance of the original $SU(N+1)$ gauge group: the symmetry breaking pattern allows for motion in the space
\begin{equation}
	 \frac{SU(N+1)}{SU(N)\times U(1)}=\mathbb{CP}(N)
\end{equation}

 The string is then endowed with an internal degree of freedom, orientational moduli that capture this phenomenon. The low-energy effective action of the string worldsheet then sees these moduli promoted to a dynamical field and produces the \cpn non-linear sigma model. 

The above construction can be generalised in the fullest extent, in that the symmetry breaking pattern of the string solution can be adjusted to produce the Flag manifold:
\begin{equation}
\frac{U(N)}{U(N_0)\times \dots U(N_p)}=\mathcal{F}_{\lbrace N_0,\dots N_p\rbrace },\quad N=\sum_{i=0}^p N_i
\end{equation}
As a special case of this construction, the Grassmannian manifold can be reached by setting $p=1$:
\begin{equation}
\frac{U(N+M)}{U(N)\times U(M)}=\mathcal{G}(N,M)
\end{equation}
The Grassmannian composite string is considerably more tractable than the generic Flag manifold string, and has undergone some analysis already, see \cite{Eto:2010aj} and a recent review \cite{Ireson:2019huc}. In many ways this text is a direct continuation of the ideas of the latter paper.

We must proceed with the following caveat. So long as the components of these composite objects all remain aligned along the same axis, the picture we outline remains valid. It is known that the full worldsheet theory for composite strings also encompasses degrees of freedom due to elementary string separation and  relative spatial orientation, as can be verified via a four-dimensional topological index computation  and a brane construction \cite{HT1}: the index found is larger than the one we will obtain in this paper.

 Attempts to take this phenomenon into account directly from field theory lead to a vastly more challenging set-up \cite{ASY}. Such analysis is however essential in order to explain the value of the 4D Witten index in this sector. An approach towards the computation of the full moduli space can be found in \cite{Eto:2005yh}, \cite{Jrev}. 

 The position of the common center of all component vortices (in the case where they do all overlap) always persists as a moduli, of course, but as always this degree of freedom decouples from colour dynamics in sufficiently supersymmetric theories. Starting from an $\mathcal{N}=2$ four dimensional theory, we expect to find a worldsheet bearing $(2,2)$ supersymmetry, in which this decoupling occurs.  In worldsheet theories of fewer supersymmetries, for instance, heterotic $(0,2)$ strings, instances are known where the fermionic components of the positional and colour zero modes mix and interact \cite{Shifman:2008wv}. Because the models we are studying are K\"{a}hler manifolds, the study of $(1,1)$-supersymmetric manifolds is trivial: the complex structure for these spaces automatically provides a SUSY enhancement to $(2,2)$. 

In this context we will show how Flag manifolds arise on the worldsheet of generic, highly-composite non-Abelian vortex strings. In Section \ref{sec4to2} we will construct an Ansatz for the fields of a particular 4D gauge theory which breaks colour symmetry in a pattern that ought to let Flag degrees of freedom appear. Then, by letting these internal degrees of freedom depend on worldsheet coordinates, we reduce the 4D action to a 2D one, corresponding to low-lying excitations of the string worldsheet. It has a very particular structure which implies gauge invariance without the existence of any tree-level dynamical gauge fields, and the structure of the coupling constants shows how different models of the same type are related to one another by a ``block merging'' phenomenon. In Section \ref{secreps} we take this worldsheet Lagrangian and re-write it in two different ways. One is a Gauged Linear Sigma Model, in which all constraints on fields are written into the action thanks to Lagrange multipliers (rather than assumed in the path integral) and gauge invariance is materialised by the introduction of an auxiliary gauge field with no kinematics of its own, obtaining a theory in which all fields are in linear representations of the symmetry groups, resembling an ordinary gauge theory. The other is a direct parametrisation of all the constraints at hand in order to obtain a true Non-Linear Sigma Model, in which the degrees of freedom exist as points on a curved manifold, i.e. a direct parametrisation of the Flag space.

These presentations of the Flag manifold Sigma Model have very recently gone under some investigation (\cite{Bykov:2019jbz},\cite{Correa:2019qox} respectively), but do not make any contact with the vortex strings which bear them, and due to this, do not bear the coupling structure derived in this work, a direct consequence of the structure of magnetic flux distribution in four dimensions, and an important tool to observe the ``block merging'' phenomenon on the worldsheet.
\section{From 4 to 2 dimensions, the structure of Composite Strings}
\label{sec4to2}

\subsection{General Composite Strings, Flag Manifolds}

We start off in four dimensional $\mathcal{N}=2$ $U(N)$ SQCD, with $N_f=N$ flavours. We introduce a Fayet-Iliopoulos $D-$term in the theory, then, the gauge symmetry becomes dynamically broken by the Higgs mechanism. The bosonic field content that interests us reduces to two gauge fields, $A_\mu$ and $A^a_\mu$ (one Abelian and the other not) as well as $N$ flavours of squarks in the fundamental representation of the gauge group, $\phi^k_A$ where $k$ and $A$ are respectively the colour and flavour indices. All other fields can be set to zero at no cost, producing a purely bosonic theory at the Bogomoln'yi point. 

The reduced Lagrangian then can be written
\begin{equation}
\mathcal{L}=\frac{1}{4g^2}\left(F^a_{\mu\nu} \right)^2 +  \frac{1}{4g^2}\left(F_{\mu\nu} \right)^2  + |D\phi|^2 +  \frac{1}{2}\tr\left(\phi^\dagger T^a \phi \right)^2 + \frac{1}{8}\left(\tr\left(\phi^\dagger \phi \right) - N\xi  \right) ^2
\end{equation}

The scalar equations of motion show that the field $\phi$ gains a diagonal VEV, enforcing a colour-flavour locked phase:  the action is invariant under leftover combined colour-flavour transformations $U(N)_{\text{diag}}$. 
\begin{equation}
	\left.\left( \phi^k_A\right) \right|_{\text{vac.}}= \sqrt{\xi}\mathbb{1}^k_A ,\quad U(N)\times U(N) \rightarrow U(N)_{\text{diag}}
\end{equation}

This pattern of symmetry breaking generates distinct topological sectors due to the following non-trivial homotopy structure
\begin{equation}
\pi_1\left( \frac{U(N)\times U(N)}{U(N)}\right) \sim \pi_1\left( U(N)\right) =\mathbb{Z}
\end{equation}
The integer that labels the equivalence classes of this homotopy is the overall winding number of a vortex. However, without breaking center symmetry, we can only prepare vortices in which all flavours have the same winding number, i.e. the string object has one unit of magnetic flux in all colour-flavours. This object has tension
\begin{equation}
	T=2\pi N \xi.
\end{equation}

While this example is simple, it seems non-minimal: the appearance of $N$ in the string tension may leave to wonder whether a object of lower tension exists, potentially by winding fewer of the gauge fields. This  can be done by breaking the center symmetry $\mathbb{Z}_N$ of $SU(N)$: of the $N$ scalar fields that exist in the model, we will assume that one of them has a topological phase factor, i.e. its phase winds around the infinite plane, while $N-1$ of them do not. This latter property implies that, unlike in the multi-Abelian string, some scalar fields remain invariant under large gauge transformations under combined motion in $U(1)$ and the center of $SU(N)$.

There are ${N}$ equivalent ways of choosing this field which experiences winding, this produces and additional selection rule due to the following non-trivial homotopy structure: dividing through by the center group,
\begin{equation}
\pi_1\left(\frac{U(1)\times SU(N)}{\mathbb{Z}_{{N}}} \right) =\mathbb{Z}_{{N}}.
\end{equation}
The objects that this construction produces, the $\mathbb{Z}_N$ string, has minimal tension \cite{ANO}
\begin{equation}
T=2\pi \xi.
\end{equation}
Because center symmetry is now broken, the flux of each individual colour-flavour is now distinguishable: strings with magnetic flux in different colour-flavours are physically identifiable, so long as we disallow residual diagonal $U(N)$ transformations. If they are allowed, they can transform one unit of flux from one colour to another. Total winding number, the topological index due to colour-flavour locking, is still conserved.

These $U(N)$ transformations are effectively new degrees of freedom for the string, specifically, moduli. By observing equivalences between these residual transformations, we can show that these moduli live in the \cpn projective space, and by enabling fluctuations of the string, this will produce the famous \cpn 2D non-linear sigma model.

From this very simple example, there are many ways by which we can make the construction more general. The first main generalisation of this process comes when one takes more than one scalar field to possess winding at infinity: we interpret this as taking several non-Abelian strings all of different colours and fusing them together. Because the objects are BPS, that is, protected not only topologically but by the conservation of certain supercharges, they exert no net force on each other, so the resulting object is stable. Assuming $L$ colours each bear one unit of magnetic flux or winding of its constituent fields, the string dynamics involve the following group quotient
 \begin{equation}
 	\frac{U(N)}{U(L)U(M)},\quad L+M=N
 \end{equation}
 which is called the Grassmannian space. Many of its properties are entirely analogous to \cpn: it is a BPS-protected object, it has finite string tension, the number of its vacua is explicitly known to be the binomial coefficient ${N \choose L}$ which can be checked by a variety of means. In addition, many important properties of the object are invariant under interchange of the numbers $L \longleftrightarrow M$.
 
 But, this is not the most generic pattern of symmetry breaking which can create these non-Abelian vortices. A further refinement involves giving different sets of colours different values of winding: we lift the requirement that the elementary non-Abelian strings we use in the process of creating our composite string all have different colours of magnetic flux. Here is a low-dimensional example of this kind of process at work: on a large circle at infinity, the scalar fields approach the following solution
 
\begin{equation}
	\Phi = \left(\begin{array}{c|cc|ccc}
	1& 0 & 0 & 0 & 0 &0  \\ 
	\hline
	0&  e^{i\theta}& 0 & 0 &0  &0  \\ 
	0&0  &  e^{i\theta}&  0&0  &0  \\ 
	\hline
	0&0  & 0 & e^{2i\theta}  & 0 &0  \\ 
	0&0  &0  & 0 & e^{2i\theta} &0  \\ 
	0 & 0 &0  & 0 &  0& e^{2i\theta}
	\end{array}  \right) 
\end{equation}
The solution breaks up into three blocks. One block of size 1 is unwound at infinity. One block of size 2 is wound once, one block of size 3 is wound twice. If the latter two had the same winding there would be no reason to consider them distinct blocks, so this extra step is necessary. This example will, following the procedure we will explain, lead to the Flag Manifold
\begin{equation}
	\mathcal{F}_{\lbrace 1,2,3\rbrace }=\frac{U(6)}{U(1)\times U(2) \times U(3)}
\end{equation}
In general, let us create $p+1$ sets of colours: we partition the total number of colours $N$ into $p+1$ integers
\begin{equation}
	N = N_0 + \dots +N_p
\end{equation}
to create collections of colours of sizes $N_0\dots N_p$. The scalar fields, inside each of these groups, will experience winding at infinity with winding numbers $q_0\dots q_p$, i.e. units of magnetic flux, all of which are different from each other. In addition, by convention we will take $q_0=0$. This is necessary to have a true non-Abelian string: much like simpler cases, the absence of winding in one direction enables some combined $U(1)$ and diagonal $SU(N)$ transformations to leave the solution invariant. Physically, a string where every colour experiences winding is simply an Abrikosov-Nielsen-Olesen (i.e. Abelian) string, we could imagine the resulting object being able to be split into an ANO string and a string of the type we have described, that is, the object we construct is``irreducible'' in that sense.

At infinity, the scalar fields in the theory tend to the following limits on a large circle:
\begin{align}
\phi^k_A =\left(\begin{array}{c|c|c|c}
\mathbb{1}_{N_0} & 0 & 0 & 0 \\
\hline
0 & e^{iq_1\theta} \mathbb{1}_{N_1} &0&0 \\
\hline
0&0& \dots &0 \\
\hline
0&0&0& e^{iq_p\theta} \mathbb{1}_{N_p}
\end{array}  \right) \label{flaglimit}
\end{align}

If the windings of the colours have this structure at infinity, we will show that strings can be constructed that respect it and that leftover $U(N)_{\text{diag.}}$ degrees of freedom active on the worldsheet of these strings exist in the following group quotient, the Flag Manifold:
\begin{equation}
\mathcal{F}_{\lbrace N_0,\dots N_p \rbrace}=\frac{U(N)}{U(N_0)\times\dots U(N_p)}
\end{equation}

We will now perform the construction of the string via the fields that compose it.
\subsection{Setting up the radial ansatz}
We propose the following Ansatz for the scalar and gauge fields. Let us label which set or collection of colours we are discussing by a generic index $\alpha$: this is not a spinorial index nor is it related to any group transformation, it is purely generational. Thus we can discuss the winding or flux number of each collection, $q_\alpha$, or their sizes $N_\alpha$. 

We introduce the total flux of the object
\begin{equation}
	Q=\sum_{\alpha=0}^{p} q_\alpha N_\alpha
\end{equation}
 with the convention that $q_0=0$.
 
 We prepare the fields in the following way: we will make use of the singular gauge description of the object, in which the winding of the scalar fields is absorbed by a change of gauge resulting in a singularity in the gauge fields themselves at the origin. To this effect we write
\begin{align}
\phi^k_A = U\left(\begin{array}{c|c|c|c}
\phi_0(r) \mathbb{1}_{N_0} & 0 & 0 & 0 \\
\hline
0 & \phi_1(r)  \mathbb{1}_{N_1} &0&0 \\
\hline
0&0& \dots &0 \\
\hline
0&0&0& \phi_p(r)  \mathbb{1}_{N_p}
\end{array}  \right) U^\dagger
\end{align}
\begin{align}
A^a_{i=1,2} T^a =\frac{1}{N} U\left( \begin{array}{c|c|c}
\left( \sum\limits_{\alpha=0}^p{q_\alpha  f_\alpha  N_\alpha}- q_0f_0  N\right)  \mathbb{1}_{N_0} & 0 & 0 \\
\hline
0&\dots &0 \\
\hline
0&0& \left( \sum\limits_{\alpha=0}^p{q_\alpha  f_\alpha  N_\alpha}- q_p  f_p  N\right)  \mathbb{1}_{N_p}
\end{array}\right) U^\dagger\,\partial_i \theta
\end{align}
\begin{equation}
A_{i=1,2} =- \frac{Q}{N}\partial_i\theta f(r) 
\end{equation}
where $U$ is an arbitrary $U(N)_{\text{diag.}}$ matrix and $\phi_{\alpha}$, $f_\alpha$, $f$ are a collection of scalar profiles  for which we specify the boundary conditions:
\begin{align}
	&\phi_\alpha(0)=0,\quad f_\alpha(0)=1,\quad f(0)=1\\
	&\phi_{\alpha}(\infty)=\sqrt{\xi},\quad f_\alpha(\infty)=0,\quad f(\infty)=0
\end{align}
With these conditions we see that the gauge fields are indeed singular at the origin since they become proportional to $\partial\theta$ but they decay to 0 at infinity. A regular gauge would require the gauge potential to be well-defined at 0 but to decay as $\frac{1}{r}$ at infinity to cancel the phase rotation due to winding of the scalar fields, in which we would see that the scalar fields do indeed tend to the limit of Eq.\ref{flaglimit}

If any two winding numbers $q_{\alpha},q_{\beta}$ are equal, two of the blocks above merge. It is therefore important that all the windings are different from each other so that the block decomposition we perform is sensible. The fact that the block decomposition changes at special points of parameter space will need to be kept in mind: it is a physical phenomenon which should be seen on the worldsheet of these strings.

 The non-Abelian part of the gauge potential is traceless as required, a fact that can be seen instantly by writing its trace as
\begin{equation}
	\sum_{\alpha,\beta=0}^{p} \left( q_\alpha f_{\alpha} - q_\beta f_{\beta}\right)N_\alpha {N_\beta}
\end{equation}
The summand is antisymmetric in $\alpha,\beta$, which leads to vanishing trace. In addition, it is clear that setting $p=1$ and $q_1=1$ reproduces the Grassmannian case.

Unlike the \cpn and Grassmannian case, the non-Abelian gauge field is now composed of several scalar functions, we introduce $p+1$ gauge profiles $f_\alpha$ though $p$ of them are actually relevant. Indeed, the profile $f_0$ is fictitious and introduced for elegance, since it always comes multiplied by the flux number $q_0=0$. Some intuition is helpful at this stage to motivate choosing to include a separate scalar profile for every block: setting all the $f_\alpha$ to be identical does not change the tracelessness of the matrix. However, there is no a-priori reason to do so since there is no symmetry principle that enforces these profiles to be equal: the $\alpha$ index is purely generational, there is not even an explicit discrete symmetry between each block. Thus, the most generic parametrisation should be used, and this will be helpful later.

With this parametrisation, the solution can be shown to preserve some supersymmetry so long as Bogomoln'yi-Prasad-Sommerfeld (BPS) first-order equations of motion are satisfied, which dictate the dynamics of the profile functions we introduced. In producing these equations, particular care needs to be taken when computing the 4D $D-$term potential, projecting it in much the same way we did the gauge scalar profiles:
\begin{equation}
D_{ij} = D \frac{1}{N} \mathbb{1}_{N} + \left(\begin{array}{c|c|c}
\left( \sum\limits_{\alpha=0}^p{q_\alpha  D_\alpha  N_\alpha}- q_0D_0  N\right)  \mathbb{1}_{N_0}  & 0 & 0  \\
\hline
0& \dots &0 \\
\hline
0&0&\left( \sum\limits_{\alpha=0}^p{q_\alpha  D_\alpha  N_\alpha}- q_p D_p  N\right)   \mathbb{1}_{N_p}
\end{array}  \right) 
\end{equation} 
Again $D_0$ is fictitious, it always comes multiplied by $q_0=0$. The peculiar shape of the $A$ and $D$ matrices is not accidental: they are the result of constructively splitting a diagonal matrix into independent trace and traceless component with the particular block-diagonal shape that we require.

When applying this decomposition to the scalar potential, the Fayet-Iliopoulos term only affects the Abelian part of the $D-$field. $D-$flatness imposes
\begin{equation}
	D=|\phi|^2 - \xi^2
\end{equation}

Once this decomposition is done, the BPS equations produce the following first-order equations of motions for the profiles we introduced

\begin{align}
& \frac{d \phi_{\beta}}{d r} -\frac{1}{N r}\left(Q f(r) -  \sum\limits_{\alpha=0}^p \left(q_\alpha f_\alpha(r)  - q_\beta f_\beta(r) \right) N_\alpha  \right)\phi_\beta = 0 \\
 -\frac{1}{r} &\dfrac{d f(r)}{dr}  + \frac{g^2}{4} \left(\sum\limits_{\alpha=0}^p N_\alpha \phi_\alpha^2 - N\xi\right) =0 \\
  -\frac{1}{r} &\dfrac{d f_\alpha(r)}{dr}  + \frac{g^2}{2 q_\alpha} \left(\phi_\alpha^2 - \phi_0^2\right) =0
\end{align}

This guarantees that the soliton is in a minimal action state: the energy density (energy per unit length) of the resulting object is then
\begin{align}
	T&= -\xi\iint F_{12}  d^2x  = -2\pi Q \xi \int_0^r f'(r) dr \nonumber\\
	T&= 2\pi Q \xi
\end{align}
This lends weight to the notion that these strings are composite objects: like in the Grassmannian string, we can view the Flag string as being the fusion of multiple elementary non-Abelian strings \cite{SYrev}, the magnetic fluxes of which sometimes align and sometimes not, as prescribed by the structure of the block sizes and relevant windings. Higher winding numbers mean more fluxes aligned with each other. Since these objects are BPS, there should be no binding energy to tie them together, and indeed we observe that the tension of the object is simply the sum of the tensions of all of its constituents. 

In order to further investigate the properties of these objects, we will require the low-energy effective action for the fluctuations of the colour degrees of freedom along this string.

\subsection{Varying the gauge moduli}
We have an arbitrary $U(N)$ degree of freedom in the string solution, the $N\times N$ matrix $U \in U(N)$. However, not all such matrices actually affect the solution. Indeed, any matrix of the form
\begin{equation}
	U= \left( \begin{array}{c|c|c|c}
	U_0 & 0 & 0 & 0 \\
	\hline
	0 & U_1&0&0 \\
	\hline
	0&0& \dots &0 \\
	\hline
	0&0&0& U_p
	\end{array} \right),\quad U_\alpha \in U(N_\alpha)
\end{equation}
does not affect the Ansatz at all, therefore we expect that the fluctuations of this parameter exist in the following quotient space
\begin{equation}
\frac{U(N)}{U(N_0)\times \dots U(N_p)}=\mathcal{F}_{\left\lbrace N_0,\, \dots\, , N_p \right\rbrace}\label{flagdef}
\end{equation}
This is the group-theoretic definition of the flag manifold.

Let us try to make explicit the degrees of freedom that should live in this Flag manifold on the string. For this purpose we break down $U$ into columns
\begin{equation}
U=(X^{(0)}\rvert \dots \rvert X^{(p)}), \quad X^{(\alpha)} = \left( X^{(\alpha)}\right) ^{A=1\dots N}_{i=1\dots N_\alpha}
\end{equation}
Each $X^{(\alpha)}$ is a rectangular $N\times N_\alpha$ matrix, a collection of columns from the square matrix $U$. The unitarity of $U$ implies the following relations among the $X$:
\begin{equation}
\left( X^{(\alpha)\dagger}\right)_{iA}\left( X^{(\beta)}\right) _{Aj} = \delta_{ij}\delta^{\alpha \beta},\quad \sum_{\alpha=0}^p X^{(\alpha)}_{Ai}X^{(\alpha)\dagger}_{iB} = \mathbb{1}_{AB}
\end{equation}

The $\alpha$ index is kept in brackets to remind ourselves no symmetry acts on it, it is purely a label or generational index. The $i$ indices range from $1$ to $N_\alpha$, strictly speaking their range is $\alpha$-dependent. Capital indices such as $A$ will range from $1$ to $N$. 

In this notation, the non-Abelian gauge field (in the singular gauge) can be written as

\begin{equation}
	A^a_{\mu=1,2} T^a_{AB}= \left( \frac{1}{N}\left(\sum\limits_{\alpha=0}^p q_\alpha N_\alpha f_\alpha(r) \right)  \mathbb{1}_{AB} - \sum_{\alpha=1}^{p} q_\alpha X^{(\alpha)}_{Ai} X^{(\alpha)\dagger}_{iB} f_\alpha (r)\right) \partial_\mu \theta 
\end{equation}
Note that $X^{(0)}$ drops out of the Ansatz.

The flag manifold \flgn as defined in Eq.(\ref{flagdef}) is a finite-dimensional space of dimension
\begin{equation}
	d= N^2 - \sum_{\alpha=0}^p N_\alpha^2
\end{equation}
Now, the $X$ variables form a unitary matrix, but not all unitary matrices acting on the string solution produce a physically different string as explained previously. This means that of the $N^2$ real degrees of freedom captured by $X$, only
\begin{equation}
	N^2 - \sum_{\alpha=0}^p N_\alpha^2 
\end{equation}
are truly physical: this is the size of the quotient in Eq.(\ref{flagdef}). We can therefore already suspect that there exists, on the worldsheet of these strings, some mechanism to remove extraneous degrees of freedom, potentially some kind of gauge invariance. However, to prove this would require producing a low-energy effective action for the worldsheet dynamics.

While we will shortly do exactly that, there is another perspective to this question which allows us to confirm our guess that the phenomenon of gauge invariance is at hand. Firstly, it occurs in the simpler cases previously studied, but in any case, intuition as to why we should expect gauge invariance to occur here comes from the linear algebraic definition of the flag manifold: a point on this manifold, a flag\footnote{A flag is thus called by analogy with a ``real world'' flag, attached to its flagpole, itself attached to the ground. It is then also, broadly speaking, a point, contained in a line, contained in a surface, contained in a volume.}, is a sequence of progressively larger hyperplanes inside $\mathbb{C}^N$
\begin{equation}
	\lbrace 0\rbrace \subset V_1 \subset V_2 \dots \subset V_p \subset \mathbb{C}^N
\end{equation} 
We specify the dimensions of these planes to be
\begin{equation}
	\left|V_1\right| = N_1,\quad \left|V_2\right| = N_1 + N_2\quad \dots \left|V_p\right| = N_1 + N_2 +\dots N_p = N- N_0
\end{equation}
Equivalently, the flag manifold can be written as a set of \textit{mutually orthogonal} (rather than progressively larger) hyperplanes 
\begin{equation}
	U_1=V_1,\quad U_2=V_2 \backslash  \ V_1 \quad, \dots U_p = V_p \backslash  (V_1 \cup V_2 \dots \cup V_{p-1}), \quad \left|U_\alpha\right| = N_\alpha
\end{equation}
When we fully specify a set of values for the $X^{(\alpha)}$ variables, we are essentially specifying an orthonormal basis for these mutually orthogonal hyperplanes $U_\alpha$. Progressively combining sets of these basis vectors together then obviously forms bases for the $V_{\alpha}$ hyperplanes, and therefore a good way of algebraically parametrising the entire space:
	\begin{align}
&\spn\left(X^{(1)} \right) = V_{1},\quad\spn\left(X^{(1)},X^{(2)} \right)  =V_{2} \dots \nonumber\\
&\spn\left(X^{(1)},\dots ,X^{(p)} \right)=V_p,\quad \spn\left(X^{(1)},\dots ,X^{(p)},X^{(0)} \right) = \mathbb{C}^N 
\end{align}

 However, this mapping is not one-to-one: many different orthonormal bases can span the same space $U_\alpha$. Two equivalent bases (spanning the same space) are related to each other by a unitary matrix inside $U(N_\alpha)$. This is a classic example of an over-representation, physically it should translate as a notion of gauge invariance. We therefore expect to see $U(N_1)\times U(N_2)\dots \times U(N_p)$ gauge invariance on the worldsheet, acting on the lowercase indices of the $X$ degrees of freedom, their column-space.

 In order to exhibit it manifestly, we must produce dynamics for the $X$ fields, and observe that global transformations can be made local. Let us assume these orientational moduli have a $\mu=0,3$ dependence. Consequently, additional gauge components need to be activated in order to preserve gauge invariance, namely $A_{0,3}$. This means additional scalar profiles for their transverse behaviour. The gauge potential needs to be complicated enough that it respects no more symmetry than the required $U(N_1)\times U(N_2)\dots \times U(N_p)$ gauge invariance, but simple enough that the scalar profiles we introduce all end up independent of each other, such that no cross-terms are generated, in order to be able to solve their equations of motion.

In order to accelerate the computation of various worldsheet components, it is convenient to use some notational shorthands for regularly-used groups of symbols.
Firstly, by insisting on the row and column nature of the objects $X^{(\alpha)}_{Ai}$ and $X^{(\alpha)\dagger}_{iA}$, keeping these lower indices in this particular ordering allows products of these objects to never be ambiguous, indices contract in a neighbour to neighbour fashion. In addition, we will be writing many different bilinears composed of $X$ variables, products thereof have particular properties that all depend on the nature of  $X^{(\alpha)}X^{\dagger(\alpha)}$ as a projector operator:
\begin{equation}
X^{(\alpha)}X^{\dagger(\alpha)}X^{(\beta)}X^{\dagger(\beta)}=X^{(\alpha)}X^{\dagger(\alpha)}\delta^{\alpha\beta}
\end{equation}
These matrices project vectors onto the orthonormal basis elements. Let us give them and their derivatives some shorthands:
\begin{equation}
P^\alpha=X^{(\alpha)}X^{\dagger(\alpha)},\quad  R^\alpha = X^{(\alpha)}\partial X^{\dagger(\alpha)},\quad L^\alpha = \partial X^{(\alpha)}X^{\dagger(\alpha)},\quad \square^\alpha = \partial X^{(\alpha)}\partial X^{\dagger(\alpha)}
\end{equation}
We avoid at all costs writing expressions where $R^0,L^0$ appear, although they will implicitly turn up in sums where their prefactor is zero, cancelling their effective contribution. 

In this notation, the scalar and gauge fields can be expressed neatly
\begin{align}
&\Phi = \sum_{\alpha=0}^p \phi_\alpha P^\alpha,\quad \partial_a\phi=\sum_{\alpha=1}^p (\phi_\alpha - \phi_0)\left(R^\alpha +L^\alpha \right) \\
&A_i = \frac{1}{N}\partial\theta\sum_{\beta=0}^{p}\left(\left(\sum_{\alpha=0}^p q_\alpha f_\alpha N_\alpha  \right) - q_\beta f_\beta N \  \right) P^\beta = \frac{1}{N}\partial\theta\sum_{\beta=0}^{p}A_\beta P^\beta \\
&\partial_a A_i = \frac{1}{N}\partial\theta\sum_{\beta=0}^{p}A_\beta\left( R^\beta + L^\beta\right) 
\end{align}

By computing worldsheet terms that exist independently of any extra gauge component $A_{0,3}$ we can intuit the form for the latter, as they should match in structure. For instance

\begin{align}
&\tr\left( \partial_a \Phi^\dagger \partial_a \Phi\right) = \sum_{\alpha=1}^p\sum_{\beta=1}^p(\phi_\alpha - \phi^0)(\phi_\beta - \phi_0)\tr\left( \left(R^\alpha +L^\alpha \right) \left(R^\beta +L^\beta \right)\right) \nonumber\\
=&2\sum_{\alpha=1}^p(\phi_\alpha - \phi_0)^2\tr\left(\square^\alpha \right) -2\sum_{\alpha=1}^p\sum_{\beta=1}^p(\phi_\alpha - \phi_0)(\phi_\beta - \phi_0) \tr\left(\square^\alpha P^\beta \right) 
\end{align}
We bring it to a more symmetric form:
\begin{align}
=&2\sum_{\alpha=1}^p(\phi_\alpha - \phi_0)^2\tr\left(\square^\alpha \right) -2\sum_{\alpha=1}^p\sum_{\beta=1}^p(\phi_\alpha - \phi_0)^2 \tr\left(\square^\alpha P^\beta \right)\nonumber \\
&-2\sum_{\alpha=1}^p\sum_{\beta=1}^p(\phi_\alpha - \phi_0)(\phi_\beta - \phi_\alpha) \tr\left(\square^\alpha P^\beta \right) \nonumber\\
=&2\sum_{\alpha=1}^p(\phi_\alpha - \phi_0)^2\tr\left(P^0\square^\alpha \right)-\sum_{\alpha=1}^p\sum_{\beta=1}^p(\phi_\alpha - \phi_\beta)(\phi_\beta - \phi_\alpha) \tr\left(\square^\alpha P^\beta \right)\nonumber\\
=&2\sum_{\alpha=1}^p\sum_{\beta=0}^{\alpha-1}(\phi_\alpha - \phi_\beta)^2\tr\left(\square^\alpha P^\beta \right)\label{flagscalar},
\end{align}
where we have used the completeness of the projection operators and the symmetry  of $\tr\left(\square^\alpha P^\beta \right)$ to isolate the symmetric part of its coefficient, and in the final expression reducing the summation range to collapse the terms in a single quantity.

This summation convention as well as the symmetries of the problem suggest we use the following prescription for $A_3$: introducing scalar profiles $\rho_{\alpha\beta}$
\begin{equation}
A_3=i\sum_{\alpha=1}^p\sum_{\beta=0}^{\alpha-1} \left(R^\alpha P^\beta - P^\beta L^\alpha \right)\rho_{\alpha\beta}(r),
\end{equation}
which can be rewritten by defining $\rho_{\alpha\beta} = \rho_{\beta\alpha}$ and using symmetries in a manifestly Hermitian form:

\begin{equation}
A_3=i\sum_{\alpha=1}^p\left(R^\beta P^0 - P^0 L^\beta \right)\rho_{0\beta} + \frac{1}{2}i\sum_{\alpha,\beta=1}^p \left(R^\beta P^\alpha - P^\alpha L^\beta \right)\rho_{\alpha\beta}(r)
\end{equation}

This substitution has two advantages: one, it reduces to the correct quantity when $p=1$ and we have a Grassmannian manifold, two, it is clear it does not accidentally respect a larger symmetry group than needed. The existence of terms $R^\alpha P^\beta$, $P^\beta L^\alpha$ forbid the $\alpha,\beta$ sectors from merging into a larger block. The first definition, with a reduced summation range, will be the one we employ the most as it can oftentimes directly enforce cancellations that would otherwise take some symmetry consideration to justify.

We add this extra gauge component to the Lagrangian and compute all contributions to the worldsheet action due to the slow fluctuations of the $X$ variables. After some tedious effort greatly hastened by our notation, the full details of which are presented in Appendix A, we obtain the following action
\begin{align}
S=\sum_{\alpha>\beta}\frac{4\pi I_{\alpha \beta} }{g^2_2}\int dt dz\,\, \tr\left(X^{(\beta)\dagger}\partial_i X^{(\alpha)} \partial_i  X^{(\alpha)^\dagger} X^{(\beta)}  \right) 
\end{align}
with a number of integration constants
\begin{align}
I_{\alpha\beta}=&\int drd\theta\,\,\left(\rho_{\alpha\beta}'^2+\frac{1}{r^2}\left( q_\alpha f_\beta-q_\beta f_\beta\right)^2\left(1-\rho_{\alpha\beta} \right)^2 + \right.\nonumber\\
&\left. \frac{\rho_{\alpha\beta}^2}{2}\left(\phi_\alpha^2+\phi_\beta^2\right) + (1-\rho_{\alpha\beta}) \left(  \phi_\alpha- \phi_\beta\right)^2\right)\label{icoef}
\end{align}
We see that the Ansatz has achieved its objective of producing a sum of surface integrals, each of which depending only on one profile $\rho_{\alpha\beta}$ at a time: no term of the form
\begin{equation}
	\sum_{\gamma}\rho_{\alpha\gamma}\rho_{\gamma \beta}
\end{equation}
occur in our result. The vanishing of these products is directly linked to the reduced number of components of the scalar profiles $\rho_{\alpha\beta}$: were it a full $(p+1)\times (p+1)$ object such terms would automatically appear and spoil the picture.

The structure of this generic integral in Eq.(\ref{icoef}) forces us to specify the boundary conditions for the $\rho$ profiles: in the singular gauge that we have chosen, $f_\alpha$ functions do not decay at zero, and since no two $q_\alpha$ windings are identical,  $1-\rho_{\alpha \beta}$  needs to vanish in order to cancel the singularity in the integral. In addition, for the soliton to be considered localised, we impose that $\rho_{\alpha \beta}$ decays at infinity. Thus,
\begin{equation}
\rho_{\alpha \beta}(0)=1,\quad \rho_{\alpha \beta}(\infty)=0.
\end{equation}

In this notation, the Grassmannian case corresponds to just one single extra profile, $\rho_{10}$ in which case the above formulae reduce correctly to previously established results, given the conventions about zero-indexed objects.

In order to find a minimal action solution, we seek to minimise the coefficients $I_{\alpha\beta}$ in addition to the worldsheet action. This produces second order equations of motion for each $\rho_{\alpha\beta}$, which we will not write. The dynamics of $\rho$ depend strongly on those of $\phi_\alpha$ and $f_\alpha$.

Quite surprisingly, for this highly supersymmetric theory, we can write an explicit solution to the equations of motion for $\rho_{\alpha\beta}$ in terms of $\phi_{\alpha,\beta}$ alone. The BPS equations will then imply the second order extremization equations that $\rho$ obeys. This fact had already been noticed in the \cpnm and Grassmannian string analysis.

In the spirit of these previous endeavours, we find that the following expression is a good solution to the equations of motion, given that the BPS equations hold:
\begin{equation}
\rho_{\alpha \beta}=\frac{\phi_{\beta} - \phi_{\alpha}}{\phi_{\beta}}
\end{equation}
This causes some tension with the boundary conditions required on the fields at hand. One case is straightforward:
\begin{equation}
\rho_{\alpha 0} = 1- \frac{\phi_{\alpha}}{\phi_0}
\end{equation}
Since $\phi_0(0)=1$ and $\phi_{\alpha>0}(0)=0$, this solution has the right boundary condition at the origin. In the \cpnm and Grassmannian case, this is enough to proceed since there is only one coefficient, $I_{10}$. In the Flag case there is a subtle issue to resolve:  $I_{\alpha\beta}$ is left undetermined when both $\alpha,\beta$ are non-zero, since both profiles in the quotient vanish at the origin. 

We take the liberty of assuming the windings $q_\alpha$ are ordered in increasing value. Then, the sums over generational indices over the worldsheet always impose $\alpha>\beta$ in our conventions.

We linearise the BPS equations around 0: for $r\ll1$,
\begin{equation}
\phi_\alpha'(r)=\frac{q_\alpha}{r}\phi_\alpha(r)\longrightarrow \phi_{\alpha}(r)\sim r^{q_\alpha}
\end{equation}
This fixes the behaviour of the $\rho$ profiles at the origin to correspond to our requirements: $\frac{\phi_\alpha}{\phi_\beta}$ correctly goes to 0 at the origin, which in turn fixes the regularity of the integral at the origin.

With this choice, it can be shown that the integration constants all simplify to the integral of a total derivative, resulting in the following expressions thanks to our well-chosen boundary conditions:
\begin{equation}
I_{\alpha\beta} = (q_\alpha - q_\beta)>0
\end{equation}
This correctly generalises the Grassmannian case, again, and like these simpler instances does not depend on the sizes of the blocks at hand, i.e. the total winding per block does not intervene.

Thus, finally, the worldsheet action for the low-energy fluctuations of this composite object is

\begin{equation}
S=\frac{4\pi }{g^2_2}\int dt dz\,\, \sum_{\alpha>\beta=0}^p(q_\alpha-q_\beta)\tr\left(X^{(\beta)\dagger}\partial_i X^{(\alpha)} \partial_i  X^{(\alpha)^\dagger} X^{(\beta)}  \right) \label{flagaction}
\end{equation}

At this point, it is worth making a number of observations about this particular action for the sigma models we derive:
\begin{itemize}
	\item Choosing the $q_\alpha$ to be strictly increasing ensures that the (Euclidean) worldsheet action is positive-definite, in particular all kinetic terms have the same sign. 
	\item Consequently, this action is minimal when the fluctuations $\partial X^{(\alpha)}$ remain in the span of $X^{(\alpha)}$: all classical solutions of the equations of motion are, at every point on the worldsheet, still a parametrisation of the Flag manifold.
	\item Generically, this flag manifold theory could have had a multitude of unrelated couplings. The computation above proves that, for the particular flag manifolds arising from the low-energy fluctuations of generic supersymmetric non-Abelian string worldsheets, the coupling constants all lock in at integer ratios of each other, since the $q_\alpha$ are all integers. 
	\item Furthermore, it is noteworthy that this quantity does not depend on any of the $N_\alpha$, only on the winding of an individual colour in the block in question. One could expect that this coefficient would depend on the total amount of flux for this block, $q_\alpha N_\alpha$, which it does not. This was also true of the Grassmannian action, which had a unit normalisation unrelated to the sizes of the gauge groups.
	\item Finally, we may observe what happens when two windings become equal: since the normalisations are proportional to differences of winding numbers, more and more parts of the action drop out completely. This is as one would expect from the 4D theory: if two winding numbers become equal, two blocks merge into one and a flag manifold with fewer inclusions appears. This can be performed all the way down to setting all the non-zero windings to be equal, in which case one recovers the Grassmannian action. We will show below the details of this phenomenon which we dub \textit{block-merging}.
\end{itemize}

Now that we have an action, we observe, as with the Grassmannian case, that the action (\ref{flagaction}) has a hidden gauge invariance. Let us act with a local symmetry transformation on the fields $X$:

\begin{equation}
	X^{(\alpha)}_{Ai}\rightarrow X^{(\alpha)}_{Ai} +  X^{(\alpha)}_{Aj}\alpha_{ji}(x) +O(\alpha^2)
\end{equation}
Then, the generic worldsheet element transforms as
\begin{align}
&\partial_a  X^{(\alpha)}_{Ai} \partial_a X^{(\alpha)\dagger}_{iB}  X^{(\beta)}_{Bj} X^{(\beta)^\dagger}_{jA}\rightarrow \partial_a  X^{(\alpha)}_{Ai} \partial_a X^{(\alpha)\dagger}_{iB}  X^{(\beta)}_{Bj} X^{(\beta)^\dagger}_{jA} \nonumber\\
&+   X^{(\alpha)}_{Ak} \alpha_{ki}(x) \partial_a X^{(\alpha)\dagger}_{iB}  X^{(\beta)}_{Bj} X^{(\beta)^\dagger}_{jA}
+\partial_a  X^{(\alpha)}_{Ai}  \alpha^\dagger_{il}(x) X^{(\alpha)\dagger}_{lB}  X^{(\beta)}_{Bj} X^{(\beta)^\dagger}_{jA} +O(\alpha^2)
\end{align}

Thanks to the orthogonality relations (still assumed imposed at the level of the partition function) the $\alpha$ dependent terms vanish identically. This proves that we have at least the gauge invariance that we require:
\begin{equation}
	U(N_1)\times U(N_2)\times\dots U(N_p)
\end{equation}
It is in fact the maximal symmetry group respected by the action above. To see this, we look at the process by which these blocks fuse. We have an enhanced symmetry if we can rewrite the action in terms of a new variable whose columns are composed of the columns inside two (or more) different $X$ variables:
\begin{equation}
	Y=\left( X^{(\alpha)}|X^{(\beta)}\right) 
\end{equation}
This object now has a column index that ranges up to $N_\alpha + N_\beta$. So long as the entire action can be rewritten in terms of $Y$ only, an identical proof as above will show that we have enhanced the gauge invariance
\begin{equation}
	U(N_\alpha)\times U(N_\beta) \rightarrow U(N_\alpha + N_\beta)
\end{equation}

However, if a term of the form
\begin{equation}
\left| \partial_a X^{(\alpha)} X^{(\beta)\dagger}\right|^2
\end{equation}
exists in the action, it is not possible to write it in terms of the merged variable.
 Only when this term is removed from the action will the enhanced symmetry occur, which is precisely controlled by the number
\begin{equation}
q_\alpha - q_\beta
\end{equation}
Thus, the winding number structure exactly controls the symmetry breaking pattern.

We can now come to a counting of the degrees of freedom in this theory and check that the result is consistent. The Flag manifold has size
\begin{equation}
	\left|\frac{U(N)}{U(N_0) \dots U(N_p)}\right| = N^2 - \sum_{\alpha=0}^{p} N_\alpha^2 = \sum_{\alpha\neq \beta=0}^p  N_\alpha N_\beta
\end{equation}
On the other hand, each field $X^{(\alpha)}$ on the worldsheet contributes $2NN_\alpha$ real degrees of freedom, of which $N_\alpha^2$ get removed by gauge invariance. Orthonormality of the entire set of the $X$ variables is representable as one large square matrix of size $\left( \sum\limits_{\alpha=1}^p N_\alpha\right)^2$. Then:
\begin{align}
N_{\text{tot.}}&=\sum_{\alpha=1}^p  2NN_\alpha - \sum_{\alpha=1}^p N_\alpha^2  -\left( \sum_{\alpha=1}^p N_\alpha\right)^2\\
&=2N(N - N_0) -  (N-N_0)^2   - \sum_{\alpha=1}^p N_{\alpha}^2\nonumber\\
&= (N+N_0)(N-N_0) - \sum_{\alpha=1}^p N_{\alpha}^2 \nonumber\\
&= N^2 - \sum_{\alpha=0}^{p} N_\alpha^2
\end{align}
All these relations are therefore crucial in the counting of degrees of freedom.

We can check that this construction does correctly reduce to the equivalent Grassmannian action as studied previously. By setting $p=1$ and $q_1=1$ we obtain
\begin{equation}
	S_{\text{Grass.}}=\frac{4\pi}{g^2}\int\,dtdz\,\tr \left| X^{(0)}\partial X^{(1)}\right|^2
\end{equation}
as required. 

While this form of the action is an efficient and clear way of representing the action, it is unpleasant to deal with due to path integral constraints imposing orthonormality relations. We seek to rewrite it in at least two different ways: the Gauged Linear Sigma Model and the usual Non-Linear Sigma model form. The former enforces gauge invariance via an auxiliary gauge field, which, if eliminated, reduces to the model we already have, the other aims to find variables which solve the constraints at the cost of living on a curved manifold. 

\section{Further Representations of the Sigma Model}
\label{secreps}

\subsection{Gauged Linear Sigma Model}
First let us focus on gauging the symmetries of the Lagrangian. We remove $X^{(0)}$ from the expressions, via the following replacement $X^{(0)}X^{(0)\dagger} = \mathbb{1} - \sum_{\alpha=1}^p X^{(\alpha)}X^{(\alpha)\dagger}$:

\begin{align}
\mathcal{L}=& \sum_{\alpha>\beta}(q_\alpha-q_\beta)\tr\left(\partial_i X^{(\alpha)} \partial_i  X^{(\alpha)^\dagger} X^{(\beta)}X^{(\beta)^\dagger } \right) \\
=&\sum_{\alpha=1}^p q_\alpha \tr\left(\mathbb{1} - \sum_{\beta=1}^p\left( X^{(\beta)}X^{(\beta)^\dagger }\right) \partial_i X^{(\alpha)} \partial_i  X^{(\alpha)^\dagger} \right)  \nonumber\\
& +  \sum_{\alpha>\beta=1}(q_\alpha-q_\beta)\tr\left(X^{(\beta)}X^{(\beta)^\dagger} \partial_i X^{(\alpha)} \partial_i  X^{(\alpha)^\dagger}  \right)\nonumber\\
=&\sum_{\alpha=1}^p q_\alpha\tr\left(\left(\mathbb{1}-  X^{(\alpha)}   X^{(\alpha)^\dagger}\right) \partial_i X^{(\alpha)} \partial_i  X^{(\alpha)^\dagger} \right) - 2\sum_{\alpha>\beta=1}^p q_\beta \tr\left(X^{(\beta)}X^{(\beta)^\dagger} \partial_i X^{(\alpha)} \partial_i  X^{(\alpha)^\dagger}  \right) \nonumber
\end{align}

In this form it can then be surmised how to form a gauge-invariant Lagrangian with an auxiliary gauge field, which would, upon integrating it out, produce the Lagrangian above. To wit, the following is satisfactory:

\begin{align}
\mathcal{L}=&\sum_{\alpha=1}^p\left(  q_\alpha \left|D_\mu X^{(\alpha)}\right|^2 + 2\sum_{\beta<\alpha}q_\beta\tr\left( i A^{(\alpha\beta)}_\mu J^{\mu(\beta\alpha)} +\text{ h.c.}\right) \right.\nonumber\\
&\left.+ \sum_{\beta,\delta<\alpha}q_\beta A^{\mu(\beta\alpha)}A^{(\alpha\delta)}_\mu X^{(\delta)\dagger}X^{(\beta)} \right) 
\end{align}
where we define the following quantities
\begin{equation}
D_\mu X^{(\alpha)} = \left(\partial_\mu X^{(\alpha)}-i  X^{(\alpha)} A^{(\alpha)}_\mu\right), \quad J^{(\alpha\beta)}_\mu = \partial_\mu X^{(\alpha)^\dagger}X^{(\beta)} = J^{(\beta\alpha)^\dagger}_\mu,\quad  A^{(\alpha\beta)}_\mu = A^{(\beta\alpha)^\dagger}_\mu
\end{equation}
The fields $A^{(\alpha)}_\mu$ is a genuine gauge field which serves to enforce $U(N_\alpha)$ gauge invariance. The currents $J^{(\alpha\beta)}_\mu$ are, despite appearances, gauge covariant quantities due to the orthogonality relations between these fields. Then, the vector fields $A^{(\alpha\beta)}_\mu$ are gauge-covariant bi-fundamental auxiliary vector fields, charged under $U(N_\alpha)\times U(N_\beta)$. They are not associated with any gauge invariance and obey $A^{(\alpha\beta)\dagger}_\mu = A^{(\beta\alpha)}_\mu$, much like $J$ does.

Now that we have placed the Lagrangian in a more usual field-theoretic form, we can exponentiate the constraints placed upon the the fields $X^{(\alpha)}$ and add them to the Lagrangian as Lagrange multipliers, to wit

\begin{align}
\mathcal{L}=&\sum_{\alpha=1}^p \left( q_\alpha \left|D_\mu X^{(\alpha)}\right|^2 + q_\alpha \tr D^{(\alpha)}\left(  X^{(\alpha)\dagger} X^{(\alpha)} - \mathbb{1}\right)  + 2\sum_{\beta<\alpha} q_\beta \left( \tr D^{(\alpha\beta)} X^{(\beta)\dagger}X^{(\alpha)}\right) \right.  \nonumber\\
&\left. + 2\sum_{\beta<\alpha}q_\beta\tr\left( i A^{(\alpha\beta)}_\mu J^{\mu(\beta\alpha)} - iA^{(\beta\alpha)}_\mu J^{\mu(\alpha\beta)}\right)+ q_\beta\sum_{\beta,\delta<\alpha} A^{\mu(\beta\alpha)}A^{(\alpha\beta)}_\mu X^{(\beta)\dagger}X^{(\beta)}\  \right) 
\end{align}
We can already notice the block merging phenomenon in this form also: whenever $q_\beta = q_\alpha$, the gauge fields $A^{(\alpha)}_\mu,\,A^{(\beta)}_\mu$ and $A^{(\alpha\beta)}_\mu$ merge into one larger gauge field, allowing for larger gauge transformations to be allowed in this action. The numbers $q_\alpha$ are not gauge couplings: a single power of $q$ multiplies terms both linear and quadratic in the auxiliary gauge fields $A^{(\alpha\beta)}$, but they determine whether these extra constraints can join into the gauge-covariant kinetic terms for the $X^{(\alpha)}$. This happens identically to the $D$ auxiliary variables. This is not accidental: this model derives from a supersymmetric theory, we expect all of these auxiliary fields to turn into components of a supermultiplet. From the form above, it is straightforward to write a SUSY action which reduces to the correct Lagrangian. 

We introduce the following superfields: $\Xi^{(\alpha)}$ a chiral multiplet bifundamental of $U(N)\times U(N_\alpha)$, $V^{(\alpha)}$ a twisted chiral multiplet in the Adjoint of $U(N_\alpha)$,  $V^{(\alpha\beta)}$ a twisted chiral multiplet in the bifundamental of $U(N_\alpha)\times U(N_\beta)$. They have the following superspace expansions
\begin{align}
	&\Xi^{(\alpha)}= X^{(\alpha)}+\theta \chi^{(\alpha)} + \theta^2 F^{(\alpha)} \nonumber \\
	&V^{(\alpha)}= \dots + \bar{\theta}\theta \sigma^{1(\alpha)} +i\theta\sigma^3\theta \sigma^{2(\alpha)}+\theta\sigma^\mu\bar{\theta} A^{(\alpha)}_\mu + \bar{\theta}^2 \theta \lambda^{(\alpha)}+\bar{\theta}^2\theta^2 D^{(\alpha)}\\
	&V^{(\alpha\beta)}= \dots + \bar{\theta}\theta \sigma^{1(\alpha\beta)}+ i \theta\sigma^3\bar{\theta}\sigma^{2(\alpha\beta)} +\theta\sigma^\mu\bar{\theta} A^{(\alpha\beta)}_\mu + \bar{\theta}^2 \theta \lambda^{(\alpha\beta)}+\bar{\theta}^2\theta^2 D^{(\alpha\beta)}\nonumber
\end{align}

We can then construct an $\mathcal{N}=(2,2)$ supersymmetric action thanks to these variables, to wit
\begin{align}
	\int d^2x \int d^2\theta d^2\bar{\theta}  &\tr\left( \sum_{\alpha=1}^p\left( q_\alpha\ \Xi^{(\alpha)}e^{V^{(\alpha)}}\Xi^{(\alpha)\dagger} \right.\right.\nonumber\\
	&\left.\left.+ 2\sum_{\beta<\alpha} q_\beta \Xi^{(\beta)}{V^{(\beta\alpha)}}\Xi^{(\alpha)^\dagger}+ \sum_{\beta,\delta<\alpha} q_\beta \Xi^{(\beta)}{V^{(\beta\alpha)}}{V^{(\alpha\delta)}}\Xi^{(\delta)^\delta}\right) \right) 
\end{align}

One may worry that $V^{(\alpha\beta)}$ does not appear in some kind of exponential in the terms above. Glossing over the technical difficulties of somehow writing a rectangular matrix in an exponential, it is in any case not necessary to do so, since they do not enforce any gauge symmetry. That is, at least, until we reach the special points where gauge symmetry is accidentally enhanced, at which point they merge with $V^{(\alpha)},\,V^{(\beta)}$ to form a larger square matrix, which can then be written as a superspace exponential to demonstrate super-gauge invariance. 

Out of superspace, this action produced is the following: we absorb $q_\alpha$ as a kinetic normalisation factor of $\Xi^{(\alpha)}$, and rescale
\begin{equation}
|\Xi|^2 \rightarrow |\Xi|^2 \frac{g^2}{4\pi} ,\quad V^{(\alpha )}\rightarrow V^{(\alpha)} \frac{4\pi}{g^2}
\end{equation}
so as to have normalised kinetic terms in the action. We obtain

\begin{align}
\mathcal{L}
&=
\tr\left(\sum_{\alpha=1}^p \left(D_\mu X^{(\alpha)} \right)^{\dagger}(D^\mu X^{(\alpha)})  - D^{(\alpha)}\left( (X^{(\alpha)\dagger} X^{(\alpha)\vphantom{\dagger}}) - 
\frac{4\pi}{g_2^2}\mathbb{1}\right) + \bar{\chi}^{(\alpha)} (\slashed{D}\chi^{(\alpha)} ) 
\right.\\
&\left.+
\left(\left( i\sqrt{2}\bar{\lambda}X^{(\alpha)}\chi^{(\alpha)}\right) +i\sqrt{2}\bar{\chi}^{(\alpha)}(\sigma^{1(\alpha)}+i\sigma^{2(\alpha)} \gamma^3)\chi^{(\alpha)}  + \text{h.c.}\right)  -2 X^{(\alpha)\dagger} (\bar{\sigma}^{(\alpha)}\sigma^{(\alpha)}) X^{(\alpha)}\right. \nonumber\\
&+\sum_{\beta<\alpha}\sqrt{\frac{q_\beta}{q_\alpha}}\left( \left( i A^{(\alpha\beta)}_\mu J^{\mu(\beta\alpha)} - iA^{(\beta\alpha)}_\mu J^{\mu(\alpha\beta)}\right)+ \sum_{\beta,\delta<\alpha} A^{\mu(\beta\alpha)}A^{(\alpha\beta)}_\mu X^{(\beta)\dagger}X^{(\beta)}\right. \nonumber\\
&\left.\left.+i\sqrt{2} \bar{\lambda}^{(\alpha\beta)X^{(\alpha)}\chi^{(\beta)}} - i\sqrt{2} \lambda^{(\beta\alpha)X^{(\beta)}\chi^{(\alpha)}} +\text{ h.c. } + 2 D^{(\alpha\beta)}X^{(\alpha)\dagger} X^{(\beta)} \right.\right.\nonumber\\
& \left. \left.+i\sqrt{2}\bar{\chi}^{(\alpha)}(\sigma^{1(\alpha\beta)}+i\sigma^{2(\alpha\beta)} \gamma^3)\chi^{(\beta)}  + \text{h.c.}  -2 X^{(\alpha)\dagger} (\bar{\sigma}^{(\alpha\beta)}\sigma^{(\alpha\beta)})X^{(\beta)}\right)\vphantom{\sqrt{\frac{a}{b}}} \right) 
\label{ws}
\end{align}
We can also introduce another representation of this action, in the form of a proper Non Linear Sigma Model, that is, using a direct parametrisation of the manifold at the cost of having a target space metric for the elementary degrees of freedom. 

\subsection{Non-Linear Sigma Model}

To perform this construction we must provide a parametrisation of the space that solves all the constraints by construction. This necessarily picks a gauge, so all the indeterminacy is lifted. We remind ourselves that the dimension of the flag manifold is
\begin{equation}
	\mathcal{F}_{\lbrace N_1\dots N_p\rbrace} = \sum_{\alpha>\beta=1}^p 2 N_\alpha N_\beta 
\end{equation}
which suggests to start by writing the fields in our previous description of the theory in the following way: we organise our degrees of freedom in the following block matrix shape

\begin{align}
	\left(Y^{(1)}|Y^{(2)}|\dots|Y^{(p)}\right) = \left( \begin{array}{c|c|c|c}
	q_1\phi_{01} & q_2 \phi_{02} & \dots & q_p \phi_{0p} \\ 
	\mathbb{1}_{N_1} & (q_2-q_1)\phi_{12} & \dots & (q_p-q_1)\phi_{1p} \\ 
	0 & \mathbb{1}_{N_2} & \dots &  \dots\\ 
	\dots & 0 & \dots &(q_{p-1} - q_p) \phi_{p-1\,\,p} \\ 
	0 & \dots &  0& \mathbb{1}_{N_p}
	\end{array} \right) 
\end{align}
where $\phi_{\beta\alpha}$ is a rectangular complex matrix with $N_\beta$ rows and $N_\alpha$ columns, and $\alpha>\beta$. We also define their complex conjugates by writing 
\begin{equation}
	\left( \phi_{\beta\alpha}\right) =\overline{\phi}_{\alpha\beta}
\end{equation}
The index structure is again representative of the row and column sizes of these rectangular blocks, allowing a check on the sanity of any products of these objects. The rectangular matrix $\phi_{\alpha\beta}$ always comes multiplied by $(q_\alpha - q_\beta)$ so that the Ansatz remains valid when the solution undergoes a block merger. Indeed, it is not merely enough that the action we are inserting this Ansatz in respects extra symmetries at certain values in parameter space, the Ansatz itself needs to obey the same property, or else it is not a good Ansatz, since it will break symmetries of the action.

 We have introduced a set of degrees of freedom in the correct number to parametrise the space in a convenient array, but this array does not (yet) satisfy the constraints in our theory, namely orthononormality.
 
First off, we ought to define the block matrix $Y^{(0)}$ to be a (non-orthonormal) basis for the complement of the space spanned by the above matrices, in a convenient notation:
\begin{equation}
	Y^{(0)}=\star \left( Y^{(1)}\wedge\dots Y^{(p)}\right) 
\end{equation}
by which we mean each individual column inside each of the blocks $Y^{(\alpha)}$ participates in this wedge product. This symbolic notion is still useful as it allows us to guess at the shape of $Y^{(0)}$ by using the usual formulae for the cross-products of vectors. As an example, let us look at the case $p=2$:
\begin{equation}
	\left( Y^{(1)}|Y^{(2)}\right) =\left( \begin{array}{cc}
	q_1 \phi_{01} & q_2 \phi_{02}\\ 
	\mathbb{1}& (q_2-q_1)\phi_{12}\\ 
	0& \mathbb{1}
	\end{array} \right),\quad Y^{(0)}=\left( \begin{array}{c}
	\mathbb{1}\\ 
	-q_1 \overline{\phi}_{10}\\ 
	q_1 (q_2-q_1) \overline{\phi}_{21}\overline{\phi}_{10} - q_2 \overline{\phi}_{20}
	\end{array} \right) 
\end{equation}
This block of columns is indeed orthogonal to the other two and its components are hinted at by the 3D cross-product formula even if strictly speaking its first component is not the matrix product of any two components of the original columns. Setting $q_2=q_1=1$ should then reduces to the Grassmannian case, providing a check of our solution: this we will do shortly.

In addition we also need to prepare from the $Y^{(\alpha)}$ vectors an orthonormal basis, to form the required $X^{(\alpha)}$ degrees of freedom. In general, orthonormal vectors are produced from a set of any linearly independent vectors via the Gram-Schmidt process. This is cumbersome to perform for block matrices: it is easy to write a normalised block vector, for instance
\begin{equation}
	\left( \begin{array}{c}
	U\\ 
	V
	\end{array} \right) \times \frac{1}{\sqrt{U^\dagger U +V^\dagger V}}
\end{equation}
but involves multiplication by an inverse square root matrix acting as its norm. These inverse square root matrices are very complicated objects in practice, in fact they are ill-defined objects: matrix square roots are defined up to a unitary matrix. In addition, being a matrix object, it rarely commutes with its surroundings, which complicates the algebra of simplifications that happen in Gram-Schmidt orthonormalisation. We present here a systematic approach to generate such a basis.

To begin, it is easy to see that the complement vector $X^{(0)}$ is defined in the following way
\begin{equation}
	Y^{(0)}=\left(\begin{array}{c}
\overline{\Delta}_{00}\\ 
	\overline{\Delta}_{10}\\ 
	\dots\\
	\overline{\Delta}_{p0}
	\end{array}  \right) ,\quad \overline{\Delta}_{\alpha 0}=\det\left( \begin{array}{ccc}
	q_1 \overline{\phi}_{10} &\mathbb{1}  & 	0\\ 
	 \dots& \dots & \mathbb{1} \\ 
 q_\alpha \overline{\phi}_{\alpha 0} &\dots  & (q_{\alpha-1}-q_\alpha) \overline{\phi}_{\alpha \,\alpha-1}
	\end{array} \right) 
\end{equation}
The ``determinant'' expressed here is not intuitively defined, if anything, because the matrix in question is not square. However, it does have the same number of row and column blocks, it is block square. The determinant operation should be thought of as indicating a rule for products between these blocks according to row expansion, resulting not in a c-number but a matrix object of size $N_\alpha\times N_0$, hence the index structure. Formally, we define this object by recursion via row expansion. To wit
\begin{equation}
	\overline{\Delta}_{00}=\mathbb{1}_{N_0},\quad \overline{\Delta}_{\alpha 0} = -\sum_{\beta<\alpha} (-1)^{\beta}(q_\alpha - q_\beta) \overline{\phi}_{\alpha\beta} \overline{\Delta}_{\beta 0}
\end{equation}
We then define the determinants $\Delta_{\alpha 0}=\left( \overline{\Delta}_{0 \alpha}\right)^\dagger$ and normalise this block by writing
\begin{equation}
	X^{(0)}= \left(\begin{array}{c}
	\mathbb{1}\\ 
	\overline{\Delta}_{10}\\ 
	\dots\\
	\overline{\Delta}_{p0}
	\end{array}  \right) \frac{1}{\sqrt{\mathbb{1}+\sum_{\beta=1}^p \Delta_{0 \beta} \overline{\Delta}_{\beta 0}}}
\end{equation}
Again, the labelling of these objects is consistent with their dimensions which allows a check at a glance of the coherence of the matrix products. In addition it is easy to check that this is directly orthogonal by construction to the $Y^{(\alpha)}$ vectors.

Furthermore, the use of these determinants allows us to express in a compact way the sought-after orthonormal basis of column blocks $X^{(\alpha)}$. This requires the introduction of yet more notational shorthands in order to be able to produce intelligible expressions. We call
\begin{equation}
	\Sigma^{(\alpha)}_{00} = \sum_{0<\beta\leq\alpha}  \Delta^{\vphantom{\dagger}}_{0\beta} \overline{\Delta}_{\beta 0},\quad \Sigma^{(0)}_{00} = \mathbb{1}
\end{equation}
such a factor already appeared in the above expression for $X^{(0)}$

It is possible to iteratively relate a certain expression involving $\Sigma$ thanks to the Sherman-Morrison formula, a specific example of the more general Woodbury identity which we will make broader use of later. To alleviate some mathematical tedium, we will consign discussions of block matrix algebra to Appendix \ref{explicitcomp}. 

An orthonormal basis of vectors spanning the relevant spaces is obtained from the coordinates defined within the $Y^{(\alpha)}$ by the following expression.

\begin{equation}
	X^{(\alpha)} = \left( \begin{array}{c}
	-\overline{\Delta}_{00} \frac{1}{\mathbb{1}+\Sigma^{(\alpha-1)}} \Delta^{\vphantom{\dagger}}_{0\alpha} \\ 
\dots	\\ 
	-\overline{\Delta}_{\alpha-1 \,0} \frac{1}{\mathbb{1}+\Sigma^{(\alpha-1)}} \Delta^{\vphantom{\dagger}}_{0\alpha} 	\\ 
	\mathbb{1}\\ 
	0\\
	\dots\\

	\end{array} \right) \frac{1}{\sqrt{\mathbb{1}+\overline{\Delta}_{\alpha 0}  \frac{1}{\mathbb{1}+\Sigma^{(\alpha -1)}_{00}}\Delta^{\vphantom{\dagger}}_{0\alpha}}}
\end{equation}
In the case of the Grassmannian manifold, the expression above does reduce to the correct result, and with a little checking it is clear that the vectors above (regardless of the normalisation factor) are all orthogonal amongst each other, are all orthogonal to $X^{(0)}$, and finally that they have unit norm. While these expressions can be derived directly from the results of Gram-Schmidt orthonormalisation, the process is tedious and unenlightening. It is enough to notice the form taken by the Gram-Schmidt solution in the case where all the $\phi_{\alpha\beta}$ are scalar to derive the form above, which then manifestly has the correct properties in the full  case.

Finally, we introduce another block matrix element, to wit 

\begin{equation}
\Gamma^{(\alpha)}_{\beta\gamma} =\left(\begin{array}{ccc}
\mathbb{1} + \overline{\Delta}_{1 0} \Delta_{01}&  \dots& \overline{\Delta}_{1 0} \Delta_{0\alpha} \\ 
\dots	& \dots & \dots \\ 
\overline{\Delta}_{\alpha 0}\Delta_{01}	&  \dots& \mathbb{1}+\overline{\Delta}_{\alpha 0} \Delta_{0\alpha}
\end{array}  \right) ^{-1}_{\beta\gamma}
\end{equation}
It is related to the previously defined object, in a way that is reminiscent of simpler identities often seen for Fubini-Study type metrics:
\begin{align}
&\Gamma^{(\alpha)} = \mathbb{1} - \overline{\Delta} \frac{1}{\mathbb{1} + \Sigma^{(\alpha)}} \Delta, \quad \frac{1}{1+\Sigma^{(\alpha)}} = \mathbb{1} - \Delta \Gamma^{(\alpha)} \overline{\Delta} \nonumber\\ 
& \Gamma^{(\alpha)}\overline{\Delta} = \overline{\Delta}\frac{1}{\mathbb{1} + \Sigma^{(\alpha)}},\quad \Delta\Gamma^{(\alpha)} = \frac{1}{\mathbb{1} + \Sigma^{(\alpha)}} \Delta
\end{align}
We introduce it to condense the expressions involved and to produce a result formally similar to the structure seen in these simpler cases. This matrix, like $\Sigma$, also obeys iterative construction laws, again given by the generic Woodbury formula.

We are now ready to write the full Non-Linear Sigma Model Action. We show the fullness of the proof of this statement in Appendix \ref{explicitcomp}. Once the substitution for the $X$ columns is performed, it can be shown that the action boils down to
\begin{equation}
	\mathcal{L} = \sum_{\alpha=1}^p (q_\alpha  -q_{\alpha+1}) \tr \left(  \frac{1}{\mathbb{1} + \Sigma^{(\alpha)}}  \partial \Delta_{0 \gamma}  \Gamma^{(\alpha)}_{\gamma\delta} \partial \overline{\Delta}_{\delta 0}\right).
\end{equation}
where, by an abuse of notation, we pick $q_{p+1} = 0$.
 
This expression is remarkable in several ways. Equating all flux numbers cancels all terms other than the very last. This last term in the summand, by itself is then action for the Grassmannian
\begin{equation}
	\frac{U(N)}{U(N_0)\times U(N_1+\dots+ N_p)}.
\end{equation}
This connects with the rigorous definition of a flag as a progressive inclusion of linear subspaces: adding the next-to-last term breaks the symmetry down to
\begin{equation}
	\frac{U(N_0+\dots +N_p)}{U(N_0) U(N_1+\dots +N_{p-1})U(N_p)}
\end{equation}
and this process carries all the way down, producing the desired Flag manifold.

 Secondly, the above action reproduces the sought-after block merger phenomenon, at least when considering merging two neighbouring blocks. Attempting to merge non-neighbouring blocks is naively incompatible with our choice of coordinates $Y^{(\alpha)}$. In any case, we have assumed from the get-go that the $q_\alpha$ windings are increasingly ordered, it is not surprising that one cannot directly see a merging of two distant blocks. It is possible to do so however: one starts by merging two neighbouring blocks of size $N_{\alpha},N_{\alpha+1}$ by setting their windings to be equal. Symmetry becomes enhanced as
\begin{equation}
U(N_\alpha)\times U(N_{\alpha+1})\rightarrow U(N_\alpha+N_{\alpha+1})
\end{equation}
At this point, we can swap over the two ``sub-blocks'' inside the newly fused block by re-ordering. Swapping them this way, then breaking the symmetry by re-introducing unequal winding, makes any \textit{specific} single degree of freedom ``travel'' to the target block to be merged with. 

Lastly, because it is built up of individual Grassmannian-like terms, it is completely straightforward to write a K\"{a}hler potential that generates this Non-Linear Sigma Model, which instantly provides us with the full $\mathcal{N}=(2,2)$ NLSM action. Flag Manifolds are known to be K\"{a}hler manifolds (in fact they are Calabi-Yau spaces, see \cite{Correa:2019qox}), but the Calabi construction for them yields one metric with no tunable parameters like we have here, thanks to our Ansatz which has this block merger property: it is rigid, where we have a deformable metric.

Let us write the K\"{a}hler potential: assuming that the field $\phi_{\alpha\beta}$ is the lowest component of an $\mathcal{N}=(2,2)$ chiral multiplet $\Phi_{\alpha,\beta}$, we write the partial determinants of these objects by recycling our notation
\begin{equation}
	\Delta_{0\beta} = \det\left( \begin{array}{ccc}
	q_1 {\Phi}_{01} &\dots  & q_\alpha {\Phi}_{0\alpha } 	\\ 
	\mathbb{1}& \dots & \dots \\ 
	0&  \mathbb{1}& (q_{\alpha-1}-q_\alpha) {\Phi}_{\alpha-1\,\alpha }
	\end{array} \right) 
\end{equation}
and
\begin{equation}
	\Sigma^{(\alpha)}_{00} =  \sum_{\beta=1}^\alpha \Delta_{0\beta} \overline{\Delta}_{\beta 0}
\end{equation}
the K\"{a}hler potential can be written
\begin{equation}
	\mathcal{K}=\sum_{\alpha=1}^p (q_\alpha - q_{\alpha+1})\tr\log\left(\mathbb{1}+ \Sigma^{(\alpha)}_{00}\right) 
\end{equation}
This reduces correctly to the Grassmannian and $\mathbb{CP}(N-1)$ cases. From this expression, it is then straightforward to define all the supermultiplet components and their interactions between them, and many geometrical insights about the theory can then be obtained. Again, performing this analysis, as one would usually an  Einstein homogeneous manifold defined as a quotient of Lie groups, would lead to a rigid K\"{a}hler potential which does not have the possibility of smoothly deforming it to manifolds with fewer degrees of freedom: we remind the reader that the flux numbers $q_\alpha$ occur not only as the leading coefficients of the terms in the Lagrangian but also in the definition of $\Delta$ itself, as reviewed above, allowing to dynamically turn on or off the required fields. This is a feature unique to our vortex construction.

For clarity, we provide in Appendix \ref{p2flag} an explicit construction of the $p=2$ flag, involving the actual physical degrees of freedom $\phi_{\alpha\beta}$, since our formulae systematically involve the determinants $\Delta_{0 \alpha}$ the formulation somewhat obscures the view.

With these algebraic details provided and the various types of actions for the model obtained, we will provide a cursory first pass over the physical properties of this class of theories.

\section{Physical Properties of the Model}
\label{secprops}

There are a few consequences that we can immediately come to. The Gauged Linear Sigma Model is particularly useful due to its similarity with ordinary gauged field theories.

\begin{figure}[h]
	\centering
	\includegraphics[width=0.4\textwidth]{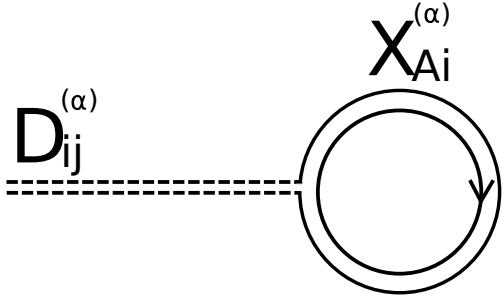}
	\caption{The one-loop diagrams leading to corrections of the FI term. They are all identical in structure, which is clear in the 't Hooft double-line prescription employed here.}\label{tadpole}
\end{figure}

Firstly, we can infer the existence of a mass gap in all of these theories. Strictly speaking, there are many couplings in the theory: every term in the sum in Eq.(\ref{flagaction}) could potentially have its own coupling, unrelated to the 4D coupling $g^2$, if not at tree level then at least as we move through the RG flow. However, the tree-level action that one derives from non-Abelian strings sees all of these couplings lock into integer ratios of each other. In addition, in the Gauged Linear Sigma Model, the coupling of the $D^{(\alpha)}$ auxiliaries to the dynamical degrees of freedom all occur identically
\begin{equation}
 \tr D^{(\alpha)} \left( X^{(\alpha)\dagger} X^{(\alpha)} - \frac{4\pi}{g^2}\mathbb{1}\right) . \label{wsymbreak}
\end{equation}
All of these FI terms could be physically different, but our construction sets them to be equal at tree level. Then, let us observe if one-loop corrections could change them. These occur due to tadpole diagrams involving loops of $X^{(\alpha)}$ as shown in Fig.(\ref{tadpole})

Clearly, all the coefficients of each $\tr D^{(\alpha)}$ term all undergo the same correction. Since they were already all equal to start with, it makes sense to say that there is one coupling for the entire theory, at one-loop order. Higher loops may spoil this picture, but given the fact that the only global symmetry in the theory is $U(N)$ and that the impact of the $q_\alpha$ windings is reduced to off-diagonal terms, it is not impossible that the theory will remain, in some form, in a ``lockstep'' phase where all the couplings to $\tr D^{(\alpha)}$ obey relations fixing them to each other, running together. 

At one-loop, therefore, the $\beta-$function for the single coupling in the theory is
\begin{equation}
	\beta(g^2) = -\frac{N}{4\pi} g^4.
\end{equation}
This immediately entails that the theory develops a mass scale, dynamically: the following mass is an RG-invariant of the theory
\begin{equation}
	\Lambda = M e^{-\frac{4\pi}{N g^2}},
\end{equation}
where $M$ is some mass parameter included in the theory through a renormalisation scheme, for instance a UV cutoff scale.

We can also comment on the number of SUSY vacua in the theory. From the four dimensional perspective, the number of distinct strings we can set up while forbidding all $U(N)$ rotations is combinatorially described by
\begin{equation}
	I_{\lbrace N_\alpha\rbrace} = \frac{N!}{N_0! \dots N_p !}.
\end{equation}
since this counts the number of ways of sprinkling the winding scalar profiles down the diagonal of the matter field Ansatz.

$U(N)$ transformations map these distinct strings onto one another, of course, and the Gauged Linear theory is a theory for the massless moduli which emerge in this picture. One way of recovering the vacua from the worldsheet theory is to make all of the fields massive, with different masses. Worldsheet masses for the $X^{(\alpha)}$ fields derive directly from four-dimensional masses for the $\Phi$ fields. Let us therefore introduce a set of masses
\begin{equation}
	m_A,\quad A\neq B \longrightarrow m_A\neq m_B
\end{equation}

At this point, the scalar potential defining the vacuum is
\begin{equation}
\sum_{\alpha} \sum_{A} 2 \left( \left(  \bar{\sigma}^{(\alpha)} - \bar{m}^A \mathbb{1} \right) \left( \sigma^{(\alpha)} - m^A\mathbb{1}\right)\right) _{ji}  X^{(\alpha)\dagger}_{iA} X^{(\alpha)}_{Aj}+ D^{(\alpha)}\left( (X^{(\alpha)\dagger} X^{(\alpha)\vphantom{\dagger}}) -
\frac{4\pi}{g_2^2}\mathbb{1}\right) 
\end{equation}
the $D-$term potential implies that the following expression is a vacuum solution
\begin{equation}
	\left(X^{(1)}|\dots|X^{(p)} \right) = \left(\begin{array}{c}
	\mathbb{1} \\ 
	0
	\end{array}  \right) ,
\end{equation}
where the upper block in the right hand matrix is of size $N_1+\dots +N_p=N-N_0$.  Since we are fully breaking $U(N)$ by introducing different masses for each flavour, we cannot map from this vacuum solution to the other ones. A generic vacuum solution, therefore, has each column in this column block above necessarily with exactly one $1$ entry, all on different rows
\begin{equation}
	X^{(\alpha)}_{A i_A} = \delta_{A i_A}
\end{equation}
There are 
\begin{equation}
	\frac{N!}{N_0! (N-N_0)!}
\end{equation}
 ways of preparing a vacuum for the $X$ fields, which is not yet correct, but we are not done constructing a vacuum solution. Indeed, the full solution also need to be compatible with the $\sigma$ part: whenever a component of $X^{(\alpha)}$ becomes non-zero, one diagonal component $\bar{\sigma}\sigma$ needs to develop a VEV in order for the relevant term to cancel. Since $N - N_0$ total columns become non-zero, all diagonal components of all $\sigma^{(\alpha)}$ fields  develop VEVs
\begin{equation}
\left( \bar{\sigma}^{(\alpha)}\sigma^{(\alpha)}\right)_{ii} = m^A\quad\text{when} \quad X^{(\alpha)}_{Ai} = 1
\end{equation}
Classically, therefore, this confirms that the theory does isolate a discrete number of vacua in the expected number. The effective potential acting on $\sigma$ can then be written, for each diagonal component $\sigma^{(\alpha)}_{ii}$,
\begin{equation}
\prod_{A=1}^N\left( \sigma^{(\alpha)}_{ii} - m^A\right) =0,
\end{equation}
i.e. the $\sigma^{(\alpha)}_{ii}$ pick out all the roots of the polynomial above. Now, from this equation, the counting of vacua can be made explicit: the ordering of the VEV components inside $\sigma^{(\alpha)}$ is irrelevant thanks to leftover $\mathbb{Z}_{N_\alpha}$ symmetry due to the Cartan generators of $U(N_\alpha)$, therefore this produces exactly 
\begin{equation}
I_{\lbrace N_\alpha\rbrace} = \frac{N!}{N_0! \dots N_p !}
\end{equation}
different solutions, from the combinatorics of picking the masses for each field $\sigma^{(\alpha)}$.

We can hypothesize that, as in the Grassmannian case, the quantum version of the equation above is simply
\begin{equation}
\prod_{A=1}^N\left( \sigma^{(\alpha)}_{ii} - m^A\right) =\Lambda^N,\label{sigmapot}
\end{equation}
this is reasonable to assume since we have a Gauged Linear Sigma Model representation, the potential for $\sigma$ likely derives from a Landau-Ginsburg effective superpotential, obtained upon integration of the full massive matter supermultiplet, in which case its quantum version proceeds from Ref.\cite{ns}. The counting is naively less obvious now, but, this equation is directly solvable if one chooses to use twisted masses:
\begin{equation}
	m^A= m\, e^{2\pi i \frac{A}{N}},\quad A=1\dots N
\end{equation}
in which case
\begin{equation}
	\sigma^{(\alpha)}_{ii} = \left|\Lambda^N + m^N\right|^{1/N} e^{2\pi i\frac{k^{(\alpha)}_i}{N}},\quad 1\leq k^{(\alpha)}_i \leq N
\end{equation}
 From this solution, we can let $m$ tend to zero, to reach the massless limit: in the quantum theory, therefore, the theory does have the correct number of vacua. A further confirmation of this property would be easily found by a direct computation of the Witten index, a topological index which is equal to the number of (unlifted) SUSY vacua. We leave this for further investigation. 

The values $k^{(\alpha)}_i$ specify the vacuum completely, but as we argued the relative orderings of these VEVs within each $\sigma^{(\alpha)}$ field are irrelevant. This means that a specific vacuum is labeled by the sets of values $K^{(\alpha)} = \left\lbrace k^{(\alpha)}_i \right \rbrace$. Defining $K^{(0)}= \mathbb{Z}_N \backslash \left(K^{(1)}\cup\dots\cup K^{(p)} \right)$, we see that the $K^{(\alpha\geq0)}$ form a partition of $\mathbb{Z}_N$ of sizes $N_0\dots N_p$, the combinatorics of which do confirm the number of vacua at hand at the quantum level. A generic vacuum state can therefore be written
 
\begin{equation}
	\left|K^{(1)},\dots,K^{(p)}\right\rangle
\end{equation}
$K^{(0)}$, being entirely determined by the other sets, does not need to figure in the quantum numbers.

By analogy with $\mathbb{CP}(N-1)$, the vacuum here has the structure of $N-N_0$ copies of $\mathbb{CP}(N-1)$, antisymmetrised and partitioned in the above way. In the case of the Grassmannian manifold, an exact identity can be written to this effect \cite{Ce-Va}: only one set of indices is needed to specify a vacuum, and the following identification is true all the way down at the quantum level
\begin{equation}
	G_{M,L} = \mathbb{CP}(N-1)^L/\mkern-6mu/ S_L.
\end{equation}
This group quotient occurring here is not an orbifold, rather, it selects the longest orbits of points under permutation of its components, i.e. it ensures that all components are all different from each other and unordered. As a result of this identification, the quantum numbers of the Grassmannian can be seen directly to be related to the structure of $\mathbb{CP}(N-1)$, since the latter has vacua labeled by $\mathbb{Z}_N$: 
\begin{equation}
\left( 	\mathbb{Z}_N\right)^L /\mkern-6mu/ S_L= \left\lbrace M\subset \mathbb{Z}_N, |M|=L \right\rbrace\label{grasslabels}
\end{equation}
The quotient operation immediately produces the quantum numbers of the vacua of the Grassmannian, subsets of ${1,\dots,N}$ of size $L$. It is not obvious how to generalise this formula, we have to produce disjoint subsets of $\mathbb{Z}_N$ of prescribed sizes with group quotients alone. The space
\begin{equation}
\left( \mathbb{CP}(N-1)^{N_1}/\mkern-6mu/ S_{N_1} \right) \times\left( \mathbb{CP}(N-N_1-1)^{N_2}/\mkern-6mu/ S_{N_2}\right)  \times \dots \left( \mathbb{CP}(N_0 + N_p-1)^{N_p}/\mkern-6mu/ S_{N_p} \right) 
\end{equation}
has the right number of elements but constructing quantum numbers like shown in Eq.(\ref{grasslabels}) from this set does not directly produce disjointed subsets of $\mathbb{Z}_{N}$. For instance, a vacuum of $\frac{U(5)}{U(2)U(2)U(1)}$ is labeled by the numbers $\left|\left\lbrace1,2\right\rbrace,\left\lbrace3,4\right\rbrace\right\rangle$, whereas the suggested group quotient construction would consider the element $\left|\left\lbrace1,2\right\rbrace,\left\lbrace1,2\right\rbrace\right\rangle$ perfectly valid: the second set in this multiplet has its range restricted from $1$ to $2$, independently of the previous set.

We propose the following representation of the space of vacua: first we define
\begin{equation}
V_0 = \emptyset
\end{equation}
then we iteratively create
\begin{align}
V_{\alpha+1} = &\left\lbrace \left( M_1,M_2, \dots,M_{\alpha+1}\right) ,\quad (M_1,\dots,M_\alpha)\in V_\alpha,\right.\nonumber \\
& \left. M_{\alpha+1} \in \left(\mathbb{Z}_N \backslash \left(M_1 \cup \dots \cup M_\alpha \right)  \right)^{N_{\alpha+1}}/\mkern-6mu/ S_{N_{\alpha+1}} \right\rbrace
\end{align}
This procedure ensures that $M_{\alpha+1}$ is one of the subsets of size $N_{\alpha+1}$ in the complement of all the previous $M_\beta$, projected duly under the symmetry group. Then, the sets of indices that generate all of the vacua of the Flag are
\begin{equation}
	\left\lbrace K_1,\dots K_p\right\rbrace \in V_p
\end{equation}By analogy we provide a construction for the entire space itself: first we define 
\begin{equation}
W_0 = \emptyset
\end{equation}
then create
\begin{align}
W_{\alpha+1} = &\left\lbrace \left( z_1,z_2, \dots,z_{\alpha+1}\right) ,\quad (z_1,\dots,z_\alpha)\in W_\alpha,\right.\nonumber \\
& \left. z_{\alpha+1} \in \left(\mathbb{CP}(N-1) \backslash \left(z_1 \cup \dots \cup z_\alpha \right)  \right)^{N_{\alpha+1}}/\mkern-6mu/ S_{N_{\alpha+1}} \right\rbrace
\end{align}
which generate the following spaces
\begin{align}	
	&W_0 = \emptyset,\quad W_1 = \frac{U(N)}{U(N_1)\times U(N-N_1)} ,\quad W_2 = \frac{U(N)}{U(N_1)\times U(N_2) \times U(N-N_1 - N_2)},\quad \nonumber\\
	&\dots,\quad  W_p = \frac{U(N)}{U(N_0)\times\dots\times U(N_p)}.
\end{align}

Finally, as previously mentioned, we anticipate that this vacuum structure derives from a Landau-Ginsburg superpotential for all the fields in $V^{(\alpha)}$, i.e. the potential written in Eq.(\ref{sigmapot}) applies not just to $\sigma^{(\alpha)}$ but to the entire multiplet. In which case, the worldsheet theory would bear kinks which interpolate between vacua. Since these states are constructed from the vacuum states of $\mathbb{CP}(N-1)$, we can already foresee, much like for the Grassmannian vacuum structure, that the spectrum of kinks of the lowest mass will interpolate between vacua with exactly one differing index. To recycle our previous concrete example of a flag, the vacua $\left| \left\lbrace1,2\right\rbrace,\left\lbrace3,4\right\rbrace\right\rangle$ and $\left|\left\lbrace1,2\right\rbrace,\left\lbrace4,5\right\rbrace\right\rangle$ will be connected by a minimal kink, but $\left|\left\lbrace1,2\right\rbrace,\left\lbrace3,4\right\rbrace\right\rangle$ and $\left|\left\lbrace1,3\right\rbrace,\left\lbrace2,4\right\rbrace\right\rangle$ will not. This analysis is occurs precisely in the Grassmannian case and the masses of all (minimal and non-minimal) kinks is known exactly thanks to the $tt^*$ equations \cite{Bourdeau:1994je} \cite{Ce-Va}, we leave the investigation for this case for a further endeavour.

\newpage
\section{Conclusions}

We have introduced the notion of a fully composite non-Abelian string: a more complex version of the Grassmannian string, it can be viewed as the admixture of several Grassmannian strings with overlapping but unequal sets of colour fluxes running through them, such that different groups of colours have different amounts of flux or winding number. The symmetry breaking that the existence of such an object enforces endows it with internal degrees of freedom, and we argued that they must exist in a Flag manifold. These spaces fully generalise the type of manifold seen previously as the target space of internal degrees of freedom in non-Abelian strings, \cpn and Grassmannian spaces. Because Grassmannian and elementary non-Abelian strings are BPS protected objects which respect some of the supersymmetry of the ambient space they exist in, we hypothetised then demonstrated that Flag strings are also BPS, and found a formula for its tension which confirms its nature as a composite object with no binding energy. Thanks to the BPS equations, we were able to write a very convenient expression for the low-energy effective action of the fluctuations of the internal degrees of freedom along the string, the Flag sigma model, in which the couplings between the various fields all depended on a single parameter, the 4D gauge coupling $g^2$, up to integer multiplicative factors related to the distribution of flux numbers $q_\alpha$ across all colours. We then computed two further presentations of this sigma model, of the Gauged Linear and Non-Linear type. The former converts all geometric constraints into very field-theoretical auxiliary terms and gauge interactions, acting on particles in linear representations of the symmetries of the system. In the end, they very much resembled Yang-Mills type theories, opening up the way for standard field-theoretical methods of analysis of this model, which we discussed. The second is the Non-Linear Sigma Model presentation, where all constraints acting on our degrees of freedom are explicitly solved for at the expense of introducing a curved target space, where quantum geometrical methods may be useful for further analysis. All three presentations of these Sigma Models, as derived from the worldsheet of vortex strings, show promising potential for future investigation.
\newpage
\section*{Acknowledgments}

We would like to thank A. Yung, M. Shifman and D. Schubring for many productive discussions. The work of E.I. is supported by the William I. Fine Theoretical Physics Institute, University of Minnesota.

\begin{appendices}
\section{Projector Algebra Notation}
We introduced the following notational shortcuts 
\begin{equation}
	P^\alpha=X^{(\alpha)}X^{\dagger(\alpha)},\quad  R^\alpha = X^{(\alpha)}\partial X^{\dagger(\alpha)},\quad L^\alpha = \partial X^{(\alpha)}X^{\dagger(\alpha)},\quad \square^\alpha = \partial X^{(\alpha)}\partial X^{\dagger(\alpha)}
\end{equation}
We avoid at all costs writing expressions where $R^0,L^0$ appear, although they will implicitly turn up in sums where their prefactor is zero, cancelling their effective contribution. 
These objects obey the following algebraic rules
\begin{align}
&P^\alpha P^\beta = P^\alpha \delta^{\alpha\beta},\quad R^\beta P^\alpha = -P^\beta L^\alpha,\nonumber \\ 
&L^\alpha R^\beta = \square^\alpha \delta^{\alpha\beta},\quad  \partial P^\alpha = (L^\alpha +R^\alpha),\nonumber\\
& P^\alpha R^\beta =\delta^{\alpha\beta}R^\alpha,\quad L^\alpha P^\beta = L^\alpha \delta^{\alpha\beta},
\end{align}
which imply the following trace identities
\begin{align}
&\tr\left(R^\alpha P^\beta R^\gamma P^\delta \right) =\delta^{\alpha\delta}\delta^{\beta\gamma} \tr\left(R^\alpha R^\beta \right)  =-\delta^{\alpha\delta}\delta^{\beta\gamma}\tr\left(\square^\alpha P^\beta\right) =-\delta^{\alpha\delta}\delta^{\beta\gamma}\tr\left(\square^\beta P^\alpha\right)\nonumber\\
=&\tr\left( P^\beta L^\alpha   P^\delta L^\gamma\right) = \delta^{\alpha\delta}\delta^{\beta\gamma} \tr\left(L^\alpha L^\beta \right) 
\end{align}

We express the 4D fields in this notation:
\begin{align}
&\Phi = \sum_{\alpha=0}^p \phi_\alpha P^\alpha,\quad \partial_a\phi=\sum_{\alpha=1}^p (\phi_\alpha - \phi_0)\left(R^\alpha +L^\alpha \right) \\
&A_i = \frac{1}{N}\partial\theta\sum_{\beta=0}^{p}\left(\left(\sum_{\alpha=0}^p q_\alpha f_\alpha N_\alpha  \right) - q_\beta f_\beta N \  \right) P^\beta = \frac{1}{N}\partial\theta\sum_{\beta=0}^{p}A_\beta P^\beta \\
&\partial_a A_i = \frac{1}{N}\partial\theta\sum_{\beta=0}^{p}A_\beta\left( R^\beta + L^\beta\right) 
\end{align}

The extra gauge components are written in the following way
\begin{equation}
A_3=i\sum_{\alpha=1}^p\sum_{\beta=0}^{\alpha-1} \left(R^\alpha P^\beta - P^\beta L^\alpha \right)\rho_{\alpha\beta}(r)
\end{equation}
Which can be rewritten by defining $\rho_{\alpha\beta} = \rho_{\beta\alpha}$ and using symmetries in a manifestly Hermitian form:
\begin{equation}
A_3=i\sum_{\alpha=1}^p\left(R^\beta P^0 - P^0 L^\beta \right)\rho_{0\beta} + \frac{1}{2}i\sum_{\alpha,\beta=1}^p \left(R^\beta P^\alpha - P^\alpha L^\beta \right)\rho_{\alpha\beta}(r)
\end{equation}

We now compute all the worldsheet contributions from the various 4D fields in the action. One was presented as motivation for the form of the new gauge components but for the sake of completeness we show all of the details.
\begin{align}
&\tr\left( \partial_a \Phi^\dagger \partial_a \Phi\right) = \sum_{\alpha=1}^p\sum_{\beta=1}^p(\phi_\alpha - \phi_0)(\phi_\beta - \phi_0)\tr\left( \left(R^\alpha +L^\alpha \right) \left(R^\beta +L^\beta \right)\right) \nonumber\\
=&2\sum_{\alpha=1}^p(\phi_\alpha - \phi_0)^2\tr\left(\square^\alpha \right) -2\sum_{\alpha=1}^p\sum_{\beta=1}^p(\phi_\alpha - \phi_0)(\phi_\beta - \phi_0) \tr\left(\square^\alpha P^\beta \right) 
\end{align}
We bring it to a more symmetric form:
\begin{align}
=&2\sum_{\alpha=1}^p(\phi_\alpha - \phi_0)^2\tr\left(\square^\alpha \right) -2\sum_{\alpha=1}^p\sum_{\beta=1}^p(\phi_\alpha - \phi_0)^2 \tr\left(\square^\alpha P^\beta \right)\nonumber \\
&-2\sum_{\alpha=1}^p\sum_{\beta=1}^p(\phi_\alpha - \phi_0)(\phi_\beta - \phi_\alpha) \tr\left(\square^\alpha P^\beta \right) \nonumber\\
=&2\sum_{\alpha=1}^p(\phi_\alpha - \phi_0)^2\tr\left(P^0\square^\alpha \right)-\sum_{\alpha=1}^p\sum_{\beta=1}^p(\phi_\alpha - \phi_\beta)(\phi_\beta - \phi_\alpha) \tr\left(\square^\alpha P^\beta \right)\nonumber\\
=&2\sum_{\alpha=1}^p\sum_{\beta=0}^{\alpha-1}(\phi_\alpha - \phi_\beta)^2\tr\left(\square^\alpha P^\beta \right)
\end{align}

\begin{align}
	&\tr\left(\partial_a A_3 \partial_a A^\dagger_3 \right) =-\sum_{\alpha=1}^p\sum_{\beta=0}^{\alpha-1}\sum_{\mu=1}^p\sum_{\nu=0}^{\mu-1} \rho_{\alpha\beta}' \rho_{\mu\nu}' \tr \left( \left(R^\alpha P^\beta - P^\beta L^\alpha \right)\left(R^\mu P^\nu - P^\mu L^\nu \right)\right) \nonumber \\
	=&-2\sum_{\alpha>\beta,\gamma>\delta}\rho_{\alpha\beta}' \rho_{\mu\nu}' \left(\tr\left( R^\alpha P^\beta R^\mu P^\nu \right) - \tr\left(P^\beta L^\alpha R^\mu P^\nu \right)  \right) \nonumber\\
		=&-2\sum_{\alpha>\beta,\gamma>\delta}\rho_{\alpha\beta}' \rho_{\mu\nu}'\tr\left( \square^\alpha P^\beta\right) \left( \delta^{\alpha\nu}\delta^{\beta\mu} -\delta^{\alpha\mu}\delta^{\beta\nu}\right) \nonumber\\
	=&2\sum_{\alpha>\beta}\rho_{\alpha\beta}'^2\tr\left( \square^\alpha P^\beta\right)	= 2\sum_{\alpha}\rho_{0\beta}'^2\tr\left( \square^\alpha P^\beta\right) + \sum_{\alpha\neq\beta = 1}^p\rho_{\alpha\beta}'^2\tr\left( \square^\alpha P^\beta\right)
\end{align}
The first term vanishes identically since it the Kronecker symbols enforce $\alpha = \nu < \mu = \beta < \alpha$, thanks to the restricted summation. In a fully summed case, a symmetry argument cancels this contribution. This matches the shape of the term generated by worldsheet variations of the scalar, without gauging. The gauging of the 4D scalars produce extra terms:

\begin{align}
	&\tr\left| A_3 \Phi\right|^2 =\sum_{\alpha>\beta}\sum_{\mu>\nu} \sum_{\lambda,\kappa=0}^p \phi_\lambda \phi_\kappa \rho_{\alpha\beta}\rho_{\mu\nu}\tr\left(\left(R^\alpha P^\beta - P^\beta L^\alpha \right)P^\lambda P^\kappa\left(P^\nu L^\mu - R^\mu P^\nu \right)  \right) \nonumber\\
		=&\sum_{\alpha>\beta}\sum_{\mu>\nu} \phi^{ 2}_\alpha \rho_{\alpha\beta}\rho_{\mu\nu}\tr\left(\left(R^\alpha P^\beta - P^\beta L^\alpha \right)\left(P^\nu L^\mu - R^\mu P^\nu \right)  \right) \\
		=&2\sum_{\alpha>\beta}  \phi^{ 2}_\alpha\rho_{\alpha\beta}^2 \tr\left(\square^\alpha P^\beta \right) = \sum_{\alpha>\beta}  \left( \phi^{ 2}_\alpha+\phi^{ 2}_\beta\right) \rho_{\alpha\beta}^2 \tr\left(\square^\alpha P^\beta \right)
\end{align}
This correctly reproduces the Grassmannian result.

\begin{align}
	&\tr\left( i \partial_a \Phi^\dagger \Phi A_3 - i A_3 \Phi^\dagger \partial_a \Phi\right) \nonumber\\
	=&\left( \sum_{\alpha>\beta} \sum_{\lambda=0}^p \sum_{\lambda=1}^p \tr \left(\left(R^\kappa+L^\kappa \right) P^\lambda \left( R^\alpha P^\beta - P^\beta L^\alpha \right)  \right)\right.\nonumber \\
	&\left.-\tr\left( \left( R^\alpha P^\beta - P^\beta L^\alpha \right)P^\lambda\left(R^\kappa+L^\kappa \right)   \right)\vphantom{\sum_{\alpha>\beta} \sum_{\lambda=0}^p \sum_{\lambda=1}^p}\right)\rho_{\alpha\beta}\phi_\lambda \left(\phi_\kappa - \phi_0\right)    \nonumber\\
	=&-2\sum_{\alpha>\beta}\rho_{\alpha\beta}\left(\phi_\alpha - \phi_\beta\right)^2\tr\left( \square^\alpha P^\beta\right) 
\end{align}

Now the components of the gauge kinetic term:

\begin{align}
&\tr\left|\partial_3 A_i\right|^2 = \frac{1}{r^2} \sum_{\alpha,\beta=1}^p \left( q_\alpha f_\alpha-\frac{1}{N} \sum_{\lambda=0}^p q_\lambda f_\lambda N_\lambda\right) \left(  q_\beta f_\beta-\frac{1}{N} \sum_{\lambda=0}^p q_\lambda f_\lambda N_\lambda\right)  \nonumber\\
&\times\tr\left(\left( L^\alpha+R^\alpha \right) \left( L^\beta+R^\beta\right) \right) \\
=&\sum_{\alpha>\beta}^p \frac{2}{r^2}\left(q_\alpha f_\alpha - q_\beta f_\beta\right)^2  \tr\left( \square^\alpha P^\beta \right) 
\end{align}
in an identical fashion as in Eq.(\ref{flagscalar})

\begin{align}
\left[A_i,A_3\right] = &\partial_i \theta\sum_{\alpha>\beta}\sum_{\lambda=0}^p \rho_{\alpha\beta} \left( q_\lambda f_\lambda - \frac{1}{N}\sum_{\kappa=0}^{p}q_\kappa f_\kappa N_\kappa\right)\nonumber\\
& \times \left( P^\lambda\left(R^\alpha P^\beta - P^\beta L^\alpha \right)- \left( R^\alpha P^\beta - P^\beta L^\alpha\right)P^\lambda \right) \nonumber\\
=&\partial_i \theta\sum_{\alpha>\beta}\rho_{\alpha\beta} \left( q_\alpha f_\beta-q_\beta f_\beta\right) \left( R^\alpha P^\beta + P^\beta L^\alpha\right) 
\end{align}
so that
\begin{align}
\left|\left[A_i,A_3\right] \right|^2= &\frac{2}{r^2}\sum_{\alpha>\beta}\rho_{\alpha\beta}^2\left( q_\lambda f_\lambda-q_\beta f_\beta\right)^2\tr\left(\square^\alpha P^\beta \right) 
\end{align}

\begin{align}
	&\tr\left(\left[A_i,A_3\right] \partial_i A_3 \right) \nonumber\\
	=&\partial_i \theta\sum_{\alpha>\beta}\sum_{\mu>\nu}\rho_{\alpha\beta} \rho_{\mu\nu}\left( q^\mu f^\mu-q^\nu f^\nu\right)\tr\left(\left(R^\alpha P^\beta + P^\beta L^\alpha \right) \left(P^\nu L^\mu - R^\mu P^\nu \right)  \right) \nonumber\\
	&=0
\end{align}
This accidental cancellation happens in the Grassmannian computation as well.
\begin{align}
&\tr\left(\left[A_i,A_3\right] \partial_3 A_i-  \partial_3 A_i\left[A_i,A_3\right]  \right) \nonumber\\
=&\frac{1}{r^2}\sum_{\alpha>\beta}\sum_{\kappa=1}^p\rho_{\alpha\beta}\left( q_\kappa f_\kappa-\frac{1}{N} \sum_{\lambda=0}^p q_\lambda f_\lambda N_\lambda\right)\left( q_\alpha f_\beta-q_\beta f_\beta\right)\nonumber\\
&\times \tr\left( \left(R^\alpha P^\beta + P^\beta L^\alpha \right) \left(R^\kappa + L^\kappa \right)+\left(R^\kappa + L^\kappa \right)\left(R^\alpha P^\beta + P^\beta L^\alpha \right) \right) \nonumber\\
=& \frac{4}{r^2}\sum_{\alpha>\beta}\rho_{\alpha\beta}\left( q_\lambda f_\lambda-q_\beta f_\beta\right)^2\tr\left(\square^\alpha P^\beta \right) 
\end{align}
Altogether this gives

\begin{align}
	S=\sum_{\alpha>\beta}\frac{4\pi I_{\alpha \beta} }{g^2_2}\int dt dz\,\, \tr\left(\partial_i X^{(\alpha)} \partial_i  X^{(\alpha)^\dagger} X^{(\beta)}X^{(\beta)^\dagger } \right) 
\end{align}
with a number of integration constants
\begin{align}
I_{\alpha\beta}=&\int drd\theta\,\,\left(\rho_{\alpha\beta}'^2+\frac{1}{r^2}\left( q_\alpha f_\beta-q_\beta f_\beta\right)^2\left(1-\rho_{\alpha\beta} \right)^2 + \right.\nonumber\\
&\left. \frac{\rho_{\alpha\beta}^2}{2}\left(\phi_\alpha^2+\phi_\beta^2\right) + (1-\rho_{\alpha\beta}) \left(  \phi_\alpha- \phi_\beta\right)^2\right)
\end{align}
 \clearpage

\section{Explicit computation of the NLSM}\label{explicitcomp}
To alleviate the discussion of the NLSM, we present in this appendix the algebraic details of the block matrix manipulations which enable us to write the NLSM action. 

We recall that we defined the Ansatz for the $X$ variables in the following way

\begin{equation}
X^{(0)}= \left(\begin{array}{c}
\mathbb{1}\\ 
\overline{\Delta}_{10}\\ 
\dots\\
\overline{\Delta}_{p0}
\end{array}  \right) \frac{1}{\sqrt{\mathbb{1}+\Sigma^{(p)}_{00}}}
\end{equation}
and
\begin{equation}
X^{(\alpha)} = \left( \begin{array}{c}
-\overline{\Delta}_{00} \frac{1}{\mathbb{1}+\Sigma^{(\alpha-1)}} \Delta^{\vphantom{\dagger}}_{0\alpha} \\ 
\dots	\\ 
-\overline{\Delta}_{\alpha-1 \,0} \frac{1}{\mathbb{1}+\Sigma^{(\alpha-1)}} \Delta^{\vphantom{\dagger}}_{0\alpha} 	\\ 
\mathbb{1}\\ 
0\\
\dots\\

\end{array} \right) \frac{1}{\sqrt{\mathbb{1}+\overline{\Delta}_{\alpha 0}  \frac{1}{\mathbb{1}+\Sigma^{(\alpha -1)}_{00}}\Delta^{\vphantom{\dagger}}_{0\alpha}}}
\end{equation}
where
\begin{equation}
\Sigma^{(\alpha)}_{00} = \sum_{0<\beta\leq\alpha}  \Delta^{\vphantom{\dagger}}_{0\beta} \overline{\Delta}_{\beta 0},\quad \Sigma^{(0)}_{00} = \mathbb{1}
\end{equation}

We immediately state the recursive relations that allow us to construct $\Sigma^{(\alpha)}_{00}$ from $\Sigma^{(\beta<\alpha)}_{00}$: it is known as the Sherman-Morrison Formula, and is a special case of the more generic Woodbury identities we will see later. All of them are simply expressions of inverses of blockwise-defined matrices in terms of the inverses of its blocks. To wit,
\begin{align}
&\frac{1}{1+\sum\limits_{\beta<\alpha}  \Delta^{\vphantom{\dagger}}_{0\beta} \overline{\Delta}_{\beta 0} + \Delta^{\vphantom{\dagger}}_{0\alpha} \overline{\Delta}_{\alpha 0}} = \frac{1}{1+\sum\limits_{\beta<\alpha}  \Delta^{\vphantom{\dagger}}_{0\beta} \overline{\Delta}_{\beta 0} } - \frac{1}{1+\sum\limits_{\beta<\alpha}  \Delta^{\vphantom{\dagger}}_{0\beta} \overline{\Delta}_{\beta 0} } \nonumber\\
&\times \Delta^{\vphantom{\dagger}}_{0\alpha} \frac{1}{\mathbb{1}+\overline{\Delta}_{\alpha 0}  \frac{1}{\mathbb{1}+\sum\limits_{\beta<\alpha}  \Delta^{\vphantom{\dagger}}_{0\beta} \overline{\Delta}_{\beta 0}}\Delta^{\vphantom{\dagger}}_{0\alpha}}\overline{\Delta}_{\alpha 0} \frac{1}{1+\sum\limits_{\beta<\alpha}  \Delta^{\vphantom{\dagger}}_{0\beta} \overline{\Delta}_{\beta 0} }\label{shermor}
\end{align}

Now, we have an ansatz for the $X$ vectors, but we will need to produce expressions for the derivatives of these vectors, and this may seem daunting. We first off mention that we will never need to introduce derivatives of the normalisation factors, since we are always projecting $\partial X^{(\beta)}$ onto a vector that $X^{(\beta)}$ is orthogonal to, either $X^{(0)}$ or another $X^{(\alpha)}$. Only derivatives of the vector-like object in the expression above needs to be differentiated. Let us proceed iteratively again, checking along the way that we recover the Grassmannian expression: firstly, we compute
\begin{equation}
q_1 \left| X^{(0)}\partial X^{(1)}\right|^2=\tr\frac{1}{\mathbb{1}+\Sigma^{(p)}_{00}}\partial\Delta^{\vphantom{\dagger}}_{01}\frac{1}{\mathbb{1}+\overline{\Delta}_{1 0} \Delta_{01}}\partial \overline{\Delta}_{1 0}
\end{equation}
If we were to set $p=1$ then we would have finished the computation and produced the NLSM. It is indeed true that the above is in that case exactly the generalised Fubini-Study metric for the Grassmannian as described in \cite{Ireson:2019huc}. It is clearer that it is the Fubini-Study metric by use of the following identity:
\begin{equation}
\frac{1}{\mathbb{1}+\overline{\Delta}_{1 0} \Delta_{01}} = \mathbb{1} - \overline{\Delta}_{1 0}\frac{1}{1+\Sigma^{(1)}_{00}}\Delta_{01} .\label{fsrel}
\end{equation}
As we add more and more terms to the Lagrangian, bigger versions of this object will occur, it will be useful to label them and write down some of their properties pre-emptively. For this purpose we previously defined 
\begin{equation}
\Gamma^{(\alpha)}_{\beta\gamma} =\left(\begin{array}{ccc}
\mathbb{1} + \overline{\Delta}_{1 0} \Delta_{01}&  \dots& \overline{\Delta}_{1 0} \Delta_{0\alpha} \\ 
\dots	& \dots & \dots \\ 
\overline{\Delta}_{\alpha 0}\Delta_{01}	&  \dots& \mathbb{1}+\overline{\Delta}_{\alpha 0} \Delta_{0\alpha}
\end{array}  \right) ^{-1}_{\beta\gamma}
\end{equation}

It is related to the previously defined object by the full version of the identities which give us Eq.(\ref{fsrel}):
\begin{align}
&\Gamma^{(\alpha)} = \mathbb{1} - \overline{\Delta} \frac{1}{\mathbb{1} + \Sigma^{(\alpha)}} \Delta, \quad \frac{1}{1+\Sigma^{(\alpha)}} = \mathbb{1} - \Delta \Gamma^{(\alpha)} \overline{\Delta} \nonumber\\ 
& \Gamma^{(\alpha)}\overline{\Delta} = \overline{\Delta}\frac{1}{\mathbb{1} + \Sigma^{(\alpha)}},\quad \Delta\Gamma^{(\alpha)} = \frac{1}{\mathbb{1} + \Sigma^{(\alpha)}} \Delta \label{fsrel2}
\end{align}

The combined use of the above equation and the ``upgrade'' identity of Eq.(\ref{shermor}) (the Sherman-Morrison identity) produces an analogous ``upgrade'' formula for $\Gamma$ matrices, called the Woodbury identity. For $\beta,\gamma<\alpha$
\begin{align}
&\Gamma^{(\alpha)}_{\alpha\alpha} = \frac{1}{\mathbb{1}+\overline{\Delta}_{\alpha 0} \frac{1}{1+\Sigma^{(\alpha-1)}} \Delta_{0 \alpha}}\nonumber\\
&\Gamma^{(\alpha)}_{\alpha\beta}=\frac{1}{\mathbb{1}+\overline{\Delta}_{\alpha 0} \frac{1}{1+\Sigma^{(\alpha-1)}} \Delta_{0 \alpha}} \overline{\Delta}_{\alpha 0} \frac{1}{\mathbb{1} + \Sigma^{(\alpha-1)}}\Delta_{0\beta}\nonumber\\
&\Gamma^{(\alpha)}_{\beta\gamma} = \Gamma^{(\alpha-1)}_{\beta\gamma} + \overline{\Delta}_{\beta 0} \frac{1}{\mathbb{1} + \Sigma^{(\alpha-1)}}\Delta_{0\alpha} \frac{1}{\mathbb{1}+\overline{\Delta}_{\alpha 0} \frac{1}{\mathbb{1}+\Sigma^{(\alpha-1)}} \Delta_{0 \alpha}}\overline{\Delta}_{\alpha 0} \frac{1}{\mathbb{1} + \Sigma^{(\alpha-1)}}\Delta_{0\gamma}
\end{align}

With these identities at hand, we can add another term to our Lagrangian. We compute
\begin{align}
&q_2 \left|X^{(0)}\partial X^{(2)}\right|^2 = q_2 \tr \left( \frac{1}{\mathbb{1} + \Sigma^{(p)}}\left(\partial \Delta_{02} \Gamma^{(2)}_{22} \partial \overline{\Delta}_{20}    + \partial \Delta_{01} \Gamma^{(2)}_{12} \partial \overline{\Delta}_{20} \right.\right. \nonumber \\
&\left.\left.  +\partial \Delta_{02} \Gamma^{(2)}_{21} \partial \overline{\Delta}_{10}+ \partial \Delta_{01} \left( \Gamma^{(2)}_{11} -\Gamma^{(1)}_{11}\right) \partial\overline{\Delta}_{10}  \right)\right) 
\end{align}

The very last term is worth noting: if we set $q_2=q_1$, a simplification occurs, and we obtain
\begin{equation}
\tr \left( \frac{1}{\mathbb{1} + \Sigma^{(p)}}\partial \Delta_{0\alpha} \Gamma^{(2)}_{\alpha\beta} \partial \overline{\Delta}_{\beta 0}  \right)  
\end{equation}
Again, if $p=2$, this is exactly the Grassmannian Sigma Model action, for the same reasons as previously. This is nothing but the block merger phenomenon at hand.

By a completely analogous iterative computation, we can write that in general
\begin{align}
&\left|X^{(0)}\partial X^{(\alpha)}\right|^2 = \tr \left( \frac{1}{\mathbb{1} + \Sigma^{(p)}}\left(\partial \Delta_{0\alpha} \Gamma^{(\alpha)}_{\alpha\alpha} \partial \overline{\Delta}_{\alpha 0}    + \partial \Delta_{0\beta} \Gamma^{(\alpha)}_{\beta\alpha} \partial \overline{\Delta}_{\alpha 0} \right.\right. \nonumber \\
&\left.\left.  +\partial \Delta_{0\alpha} \Gamma^{(\alpha)}_{\alpha\gamma} \partial \overline{\Delta}_{\gamma 0}+ \partial \Delta_{0\beta} \left( \Gamma^{(\alpha)}_{\beta \gamma} -\Gamma^{(\alpha - 1)}_{\beta\gamma}\right) \partial\overline{\Delta}_{\gamma 0}  \right)\right) \nonumber\\
& =\tr\left(  \frac{1}{\mathbb{1} + \Sigma^{(p)}} \partial \Delta_{0 \beta} \Gamma^{(\alpha)}_{\beta\gamma} \partial \overline{\Delta}_{\gamma 0}  -  \frac{1}{\mathbb{1} + \Sigma^{(p)}} \partial \Delta_{0 \beta} \Gamma^{(\alpha-1)}_{\beta\gamma} \partial \overline{\Delta}_{\gamma 0}\right)
\end{align}

We are not yet finished however as the expression above is only part of the full answer. We must also compute the terms involving different $X^{(\alpha>0)}$. The computational techniques to do so are all identical to the ones already seen, we thus get, for $\alpha>\beta$

\begin{align}
&(q^\alpha - q^\beta) \left| X^{(\alpha)} \partial X^{(\beta)} \right|^2 =\nonumber\\
& 	(q^\alpha - q^\beta) \tr \left( \left( \frac{1}{\mathbb{1} + \Sigma^{(\alpha)}} - \frac{1}{\mathbb{1} + \Sigma^{(\alpha -1)}}\right) \left( \partial \Delta_{0 \gamma}\right) \left( \Gamma^{(\beta)}_{\gamma\delta} - \Gamma^{(\beta-1)}_{\gamma\delta}\right) \left(\partial \overline{\Delta}_{\delta 0} \right) \right) 
\end{align}

When all terms are added to the Lagrangian, many cross-simplifications occur due to repeated, cancelling terms in the expressions. Let us, for simplicity, define for $\alpha>\beta$
\begin{equation}
G_{\alpha,\beta} = \tr \left(  \frac{1}{\mathbb{1} + \Sigma^{(\alpha)}}  \partial \Delta_{0 \gamma}  \Gamma^{(\beta)}_{\gamma\delta} \partial \overline{\Delta}_{\delta 0}\right)
\end{equation}

It is then simple enough to extract which of the Lagrangian terms contributes an individual $G$ term, and sum up all the flux numbers to obtain its leading coefficient. This splits up into several cases, since the cases where one of the indices is $p$ is special. We obtain the following terms, for $\beta-1<\alpha-1<p$:

\begin{align}
&G_{p,p}\times q_p \\
&G_{p,p-1}\times\left( -q_p +q_{p-1} + (q_p - q_{p-1})\right)=0 \\
&G_{p,\alpha}\times\left( q_\alpha - q_{\alpha+1} + (q_p - q_\alpha) - (q_p - q_{\alpha+1})\right)=0 \\
&G_{\alpha,\alpha} = (q_\alpha - q_{\alpha+1}) \\
&G_{\alpha,\alpha-1} = (q_\alpha - q_{\alpha-1}) - (q_{\alpha+1} - q_{\alpha-1})  + (q_{\alpha+1} -q_{\alpha})=0\\
&G_{\alpha,\beta} = (q_\alpha-q_\beta) - (q_{\alpha+1} - q_\beta)- (q_{\alpha} - q_{\beta+1})+ (q_{\alpha+1} - q_{\beta+1})=0
\end{align}
By an abuse of notation we can write $q_{p+1}=0$ to have the total Lagrangian be
\begin{equation}
\mathcal{L} = \sum_{\alpha=1}^p (q_\alpha  -q_{\alpha+1}) \tr \left(  \frac{1}{\mathbb{1} + \Sigma^{(\alpha)}}  \partial \Delta_{0 \gamma}  \Gamma^{(\alpha)}_{\gamma\delta} \partial \overline{\Delta}_{\delta 0}\right).
\end{equation}

\section{The p=2 Flag Manifold NLSM written in block components}\label{p2flag}

We will here show the explicit construction of the simplest type of Flag manifold, when $p=2$: by this we mean fully substituting the $\Delta$ determinants for the actual elementary degrees of freedom, the $\phi$ fields, as the generic formula obscures the view of their input in the model. 

 A few intermediary additional simplifications occur in the case of $\frac{U(3)}{U(1)^3}$ i.e. when $p=2$ and all variables are true scalars, not matrices, in being able to commute terms past each other. We will write the final form of the action in a way that makes it clearly analogous to the Fubini-Study metric, in which these extra simplifications are unneeded.

We define the variables $\phi_{\alpha,\beta}$ in the following way
\begin{equation}
\left( Y^{(1)}|Y^{(2)}\right) =\left( \begin{array}{cc}
\phi_{01} & \phi_{02} \\ 
\mathbb{1} & \phi_{12} \\ 
0& \mathbb{1}
\end{array} \right)
\end{equation}
We suppress factors of $q_\alpha$ in this definition to keep expressions tidy but they are otherwise necessary in order for this Ansatz to reduce correctly. Notably we recall that $\phi_{12}$ has a factor of $q_1 -q_2$ leading it, thus causing it to drop out of the solution altogether at the special point, when the space becomes a Grassmannian.

Firstly we create the partial determinants $\Delta_{0\alpha}$:
\begin{align}
	&\Delta_{01}=-\phi_{01}\\
	&\Delta_{02}=\phi_{01}\phi_{12} - \phi_{02}
\end{align}

We then define the objects $\Sigma^{(\alpha)}_{00}$:
\begin{align}
&\Sigma^{(1)}_{00}=\phi_{01} \overline{\phi}_{10}\\
&\Sigma^{(2)}_{00}=\phi_{01} \overline{\phi}_{10} + \left(\phi_{02} -\phi_{01}\phi_{12}  \right) \left( \overline{\phi}_{20} - \overline{\phi}_{10} \right) 
\end{align}

We now define
\begin{equation}
	\Gamma^{(\alpha)} = \left(\begin{array}{cc}
	\mathbb{1}+ \overline{\phi}_{10}\phi_{01}& \overline{\phi}_{10}\phi_{02} + \phi_{12}  \\ 
\overline{\phi}_{21} + \overline{\phi}_{20}\phi_{01}	& \mathbb{1} + \overline{\phi}_{20}\phi_{02} + \overline{\phi}_{21}\phi_{12} 
	\end{array}  \right)^{-1}
\end{equation}

Symbolically this definition is practical to keep expressions tidy, but it is difficult to express in components: repeated use of the Woodbury formula is required to define it explicitly. Instead we will define this object through its relation to $\Sigma^{(\alpha)}$, emphasizing the connection to the Fubini-Study metric:

\begin{align}
&\Gamma^{(1)}=\mathbb{1} - \overline{\phi}_{10} \frac{1}{\mathbb{1}+\phi_{01} \overline{\phi}_{10}}\phi_{01}\\
&\Gamma^{(2)}=\left( \begin{array}{cc}
\mathbb{1} & 0 \\ 
0 & \mathbb{1}
\end{array} \right) \\
& -\left(\begin{array}{cc}
\overline{\phi}_{10} \\ 
\overline{\phi}_{20} - \overline{\phi}_{21}\overline{\phi}_{10}
\end{array}  \right) 
\frac{1}{\mathbb{1}+\phi_{01} \overline{\phi}_{10} + \left(\phi_{02} -\phi_{01}\phi_{12}  \right) \left( \overline{\phi}_{20} - \overline{\phi}_{21}\overline{\phi}_{10} \right)}\left( \begin{array}{cc}
\phi_{01} & \phi_{02} -\phi_{01}\phi_{12}
\end{array} \right) \nonumber
\end{align}

From there we define $X^{(0)},\,X^{(1)},\,X^{(2)}$:
\begin{align}
&X^{(0)} = \left( \begin{array}{c}
\mathbb{1} \\ 
-\overline{\phi}_{10} \\ 
-\overline{\phi}_{20} +\overline{\phi}_{21}\overline{\phi}_{10}
\end{array} \right) \frac{1}{\sqrt{\mathbb{1}+\phi_{01}\overline{\phi}_{10}+\left({\phi}_{02} +{\phi}_{01}{\phi}_{12} \right) \left( \overline{\phi}_{20} +\overline{\phi}_{21}\overline{\phi}_{10}\right) }}\\
&X^{(1)} = \left( \begin{array}{c}
\phi_{01} \\ 
\mathbb{1}\\
0
\end{array} \right) \frac{1}{\sqrt{\mathbb{1}+\overline{\phi}_{10}\phi_{01}}}\\
& X^{(2)} =\left( \left( \begin{array}{c}
\frac{1}{\mathbb{1}+\overline{\phi}_{10}\phi_{01}}\left(\phi_{02}-\phi_{01}\phi_{12} \right) \\
\overline{\phi}_{10}\frac{1}{\mathbb{1}+\overline{\phi}_{10}\phi_{01}}\left(\phi_{01}\phi_{12} - \phi_{02} \right) \\
\mathbb{1}
\end{array} \right) \nonumber\right.\\
&\times \left. \frac{1}{\sqrt{ \mathbb{1}+\left( \overline{\phi}_{20} - \overline{\phi}_{21}\overline{\phi}_{10}\right)\frac{1}{\mathbb{1}+\phi_{01} \overline{\phi}_{10}} \left(\phi_{02}-\phi_{01}\phi_{12} \right)  }} \right)
\end{align}

We can then compose the full Lagrangian.

\begin{align}
&\mathcal{L}=(q_1-q_2)\tr\left( \frac{1}{{\mathbb{1}+\phi_{01}\overline{\phi}_{10}}}\partial \phi_{01}  \left( \mathbb{1} - \overline{\phi}_{10} \frac{1}{\mathbb{1}+\phi_{01} \overline{\phi}_{10}}\phi_{01}\right)\partial \overline{\phi}_{10} \right) \nonumber\\
&+q_2\tr\left( \frac{1}{\mathbb{1}+\phi_{01} \overline{\phi}_{10} + \left(\phi_{02} -\phi_{01}\phi_{12}  \right) \left( \overline{\phi}_{20} - \overline{\phi}_{21}\overline{\phi}_{10} \right)}\left( \begin{array}{cc}
\partial\phi_{01} & \partial\left(\phi_{02} -\phi_{01}\phi_{12}\right)
\end{array} \right) \right.\nonumber\\
&\left.\left[\left( \begin{array}{cc}
\mathbb{1} & 0 \\ 
0 & \mathbb{1}
\end{array} \right)  -\left(\begin{array}{c}
\overline{\phi}_{10} \\ 
\overline{\phi}_{20} - \overline{\phi}_{21}\overline{\phi}_{10}
\end{array}  \right) 
 \frac{1}{\mathbb{1}+\phi_{01} \overline{\phi}_{10} + \left(\phi_{02} -\phi_{01}\phi_{12}  \right) \left( \overline{\phi}_{20} - \overline{\phi}_{21}\overline{\phi}_{10} \right)}\right.\right.\nonumber\\
&\left.\left.\left( \begin{array}{cc}
\phi_{01} & \phi_{02} -\phi_{01}\phi_{12}
\end{array} \right)\right]\left(\begin{array}{c}
\partial\overline{\phi}_{10} \\ 
\partial\left( \overline{\phi}_{20} - \overline{\phi}_{21}\overline{\phi}_{10}\right) 
\end{array}  \right) \right)
\end{align}

Setting $q_1 = q_2$ in practice also cancels all contributions from $\phi_{12}$ since we have suppressed some prefactors in the above expression. Performing these cancellations we see that the action reduces to the usual Fubini-Study metric of a Grassmannian: the first term cancels altogether and the second, thanks to our substitution for $\Gamma^{(2)}$, is already in the tell-tale shape that the action is often presented in.

\end{appendices}

\end{document}